\documentclass{article}

\usepackage{arxiv}

\usepackage[utf8]{inputenc} % allow utf-8 input
\usepackage[T1]{fontenc}    % use 8-bit T1 fonts
\usepackage{hyperref}       % hyperlinks
\usepackage{url}            % simple URL typesetting
\usepackage{booktabs}       % professional-quality tables
\usepackage{amsfonts}       % blackboard math symbols
\usepackage{nicefrac}       % compact symbols for 1/2, etc.
\usepackage{microtype}      % microtypography
\usepackage{multirow}
\usepackage{lipsum}
\usepackage{graphicx}
\usepackage{amsmath}
\usepackage{nccmath}
\usepackage[super]{nth}
\usepackage{float}
\graphicspath{ {./images/} }

\title{Unveiling Spatial Patterns of Disaster Impacts and Recovery Using Credit Card Transaction Variances}

\author{
 Faxi Yuan \\
  Urban Resilience.AI Lab\\
  Zachry Department of Civil and \\Environmental Engineering\\
  Texas A\&M University\\
  College Station, TX 77843 \\
  \texttt{faxi.yuan@tamu.edu} \\
  \And
 Amir Esmalian \\
  Urban Resilience.AI Lab\\
  Zachry Department of Civil and \\Environmental Engineering\\
  Texas A\&M University\\
  College Station, TX 77843 \\
  \texttt{amiresmalian@tamu.edu} \\
  \And
 Bora Oztekin \\
  Urban Resilience.AI Lab\\
  Department of Computer Science and Engineering\\
  Texas A\&M University\\
  College Station, TX 77843 \\
  \texttt{bora@tamu.edu} \\
  \And
 Ali Mostafavi \\
  Urban Resilience.AI Lab\\
  Zachry Department of Civil and \\Environmental Engineering\\
  Texas A\&M University\\
  College Station, TX 77843 \\
  \texttt{amostafavi@civil.tamu.edu} \\
}

\begin{document}
\maketitle
\begin{abstract}
The objective of this study is to examine spatial patterns of disaster impacts and recovery of communities based on variances in credit card transactions. Such variances could capture the collective effects of household impacts, disrupted accesses, and business closures, and thus provide an integrative measure for examining disaster impacts and community recovery in disasters. Existing studies depend mainly on survey and sociodemographic data for disaster impacts and recovery effort evaluations, although such data has limitations, including large data collection efforts and delayed timeliness results. In addition, there are very few studies have concentrated on spatial patterns and disparities of disaster impacts and short-term recovery of communities, although such investigation can enhance situational awareness during disasters and support the identification of disparate spatial patterns of disaster impacts and recovery in the impacted regions. This study examines credit card transaction data Harris County (Texas, USA) during Hurricane Harvey in 2017 to explore spatial patterns of disaster impacts and recovery during from the perspective of community residents and businesses at ZIP code and county scales, respectively, and to further investigate their spatial disparities across ZIP codes. The results indicate that individuals in ZIP codes with populations of higher income experienced more severe disaster impact and recovered more quickly than those located in lower-income ZIP codes for most business sectors. Our findings not only enhance the understanding of spatial patterns and disparities in disaster impacts and recovery for better community resilience assessment, but also could benefit emergency managers, city planners, and public officials in harnessing population activity data, using credit card transactions as a proxy for activity, to improve situational awareness and resource allocation.
\end{abstract}

% keywords can be removed

\section{Introduction}
The objective of this study is to examine spatial patterns of disaster impacts and recovery of communities based on variances in credit card transactions. Such variances could capture the collective effects of household impacts, disrupted accesses, and business closures, and thus provide an integrative measure for examining disaster impacts and community recovery in disasters. Examining community resilience, including disaster impacts and recovery duration (1), from a systems perspective calls for the inclusion of interactions among system components (2–5). Community systems include businesses which provide the products and services, residents who use the products and services, and infrastructure (e.g., roads) that provides access to businesses (6). Perturbations caused by natural hazards (e.g., hurricanes and flooding) could impact residents, businesses, and infrastructure systems. The collective effects of these impacts could be captured using credit card transactions as a proxy for population activity data. 

Understanding the state of the community, including immediate disaster impacts and short-term recovery duration, provides the basis for resource allocation and for prioritization of recovery strategies (7-9). Short-term recovery lasts from days to weeks and includes assessment of the damages and needs, restoration of basic infrastructure, and mobilization of recovery resources (10). Current approaches for assessing disaster impacts and recovery of communities, however, do not consider the dynamic interactions among the systems and their components. The primary tool that researchers have implemented in examining community resilience is collection of survey data (11–14). While surveys are helpful approaches for exploring determinants of community impact and recovery and provide quantitative data, this approach suffers certain limitations in the examination of community resilience (15). First, it may not be feasible to collect large and representative samples, as implementation of surveys requires extensive time and resources. Second, surveys often fail to provide information about the complex interactions underlying impacts and recovery.

Understanding the spatial patterns and disparities related to impacts and short-term recovery across various regions not only enhances situational awareness during disasters but also supports the identification of regions with more severe disaster impacts and slower recovery (16–18). Different regions in a community may not experience the same disturbance impacts (21,22). The disparate impacts and recovery patterns from the disasters are not limited to the unequal exposure of the areas to the hazard (12,23). The condition of the infrastructure, state of business, and demographic characteristics all play a role in the extent to which certain areas are impacted and could recover from the disturbances. With the growing availability of population activity data, it is now possible for researchers to explore such spatiotemporal aspects and further resolve the previous knowledge gap in quantifying community resilience in terms of disaster impacts and short-term recovery duration (24,25). Population activity data (e.g., mobility, visits to points of interest, and credit card transactions) provide researchers with fundamental human activity information (26). Researchers have utilized these types of data to explore and study different aspects of disasters (27–29). The current applications of population activity data in the context of disasters includes examining evacuation patterns (30), understanding social impacts (28,31), evaluating economic resilience (32,33), and assessing business recovery (34). Population activity data could effectively capture interactions of residents with physical systems, such as businesses and infrastructures. Nevertheless, limited studies using population activity data to explore disaster impacts and short-term recovery exist. 

In this study, we use Safegraph’s credit card transaction (CCT) data as an aggregated measure to examine the state of the community in terms of disaster impacts and recovery duration (Fig. 1). The CCT data include information about ZIP codes where individual residents live and their credit card transaction history, which provides information regarding business categories and transaction dates to enable determination of variances in credit card transactions prior to, during, and in the aftermath of disasters. This data is captured and evaluated to find the spatiotemporal patterns. Perturbations caused by disasters could affect households, infrastructure systems that provide access to business, as well as businesses themselves. The collective effects of these disruptions could be captured in the fluctuations of credit card transactions. For example, if households are economically impacted by the disaster, or if they could not access businesses because of road inundation/closure, or if businesses are closed due to damage, the effects of these perturbations are reflected in the credit card transactions, as illustrated in Fig. 1. In this study, we employ the CCT data to unveil the spatial patterns and disparities of disaster impacts and recovery duration from the perspective of community residents across the ZIP codes in the context of Hurricane Harvey which made landfall in August 2017 in Harris County, Texas. In addition, we explore the disparities in disaster impacts and recovery duration across various business sectors at the county level from a business perspective. 

\begin{figure}[!ht]
\centering
\includegraphics[width=0.85\linewidth]{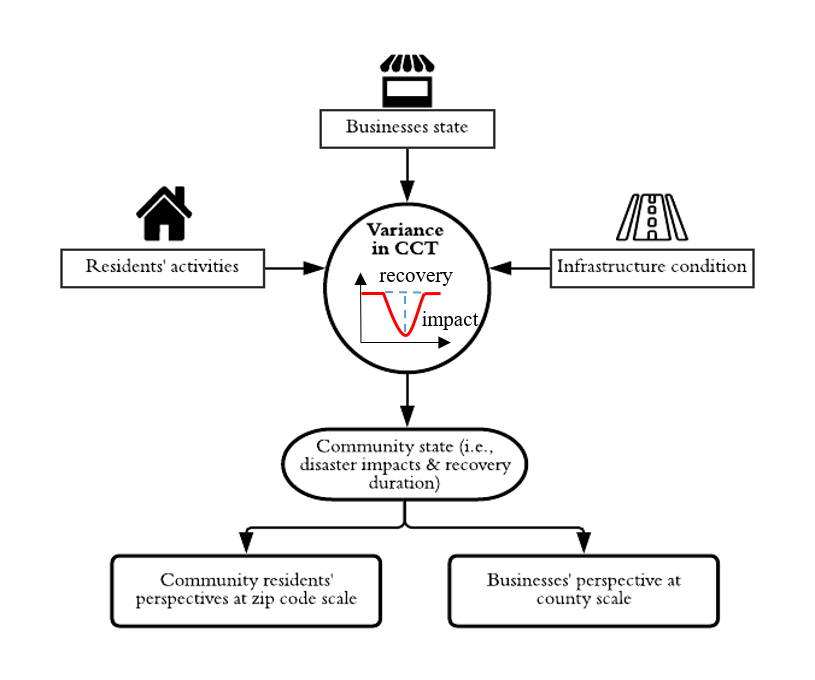}
\caption{Credit card transaction data for revealing disparities in impact and recovery of affected communities.}
\end{figure}

\section{Results}

In this section, we first describe calculation results for values of disaster impacts and recovery duration across business sectors and ZIP codes to reveal their spatial patterns across Harris County. Second, we discuss the results of the stepwise regression model to explore spatial disparities in disaster impacts and recovery duration across different ZIP codes with different flood claims and sociodemographic attributes. Finally, we present the results related to the disaster impacts and recovery duration for all the defined business sectors at the county level. Results are presented in plots related to three sectors (drugstore, health care, and utilities–electric, gas, water, and sanitary) of the defined business sectors as examples of how we quantified disaster impacts and recovery duration with the CCT variance data. Results related to the other 18 business sectors are presented in the Supplementary Information. 

\subsection{Spatial pattern analysis results} 
\subsubsection{\textit{Geographic distribution of disaster impact and recovery duration}}
We computed the community state indices (i.e., disaster impacts and recovery duration) using Eq. (1-4) for business sectors for 142 ZIP codes in Harris County. Using ArcGIS, geographic distributions of these two indices for the grocery sector as an example are presented in Fig. 2. For the grocery sector, Fig. 2a shows disaster impacts; Fig. 2b shows recovery duration. In both figures, oval 1 indicates clusters of similar disaster impact level; oval 2 indicates clusters of dissimilar recovery duration levels. To better capture the spatial patterns of these two indices, we describe the spatial autocorrelation analysis results in the following sections. 

\begin{figure}[ht]
\centering
\includegraphics[width=\linewidth]{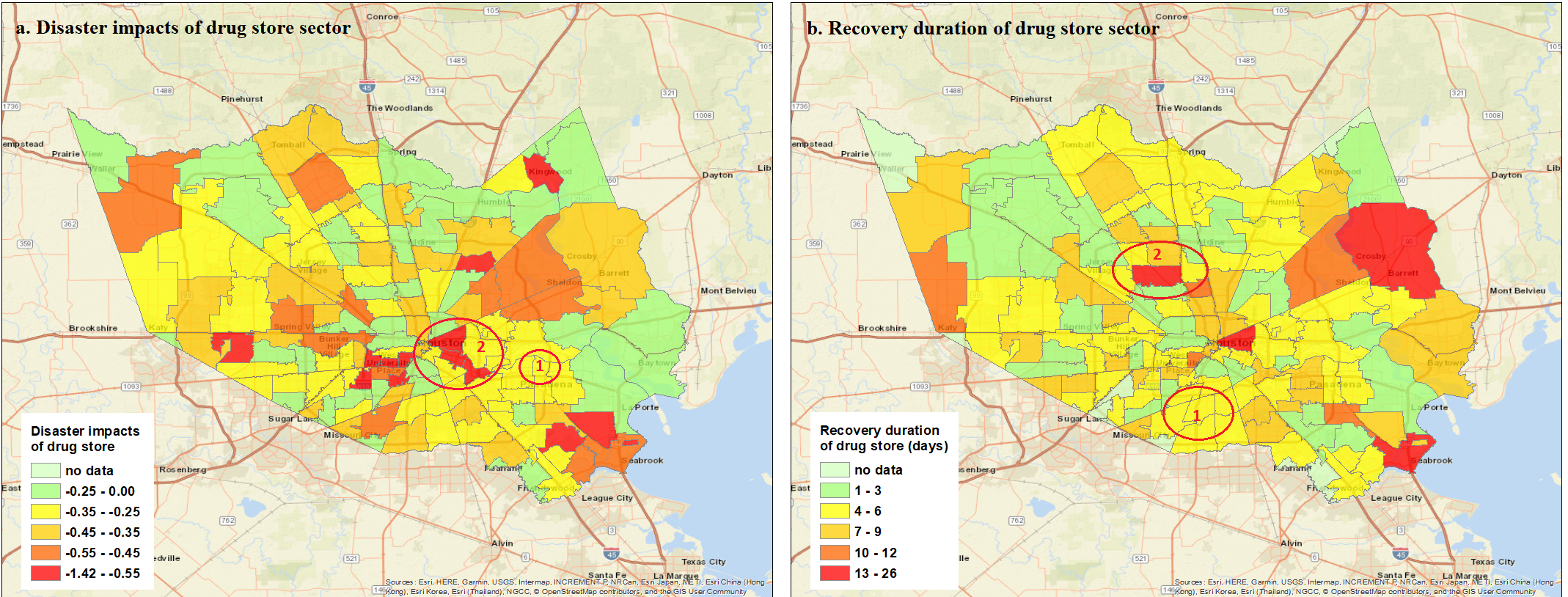}
\caption{Geographic spatial distribution for the grocery sector: a, disaster impact (left); b, recovery duration (right). a, b, Oval 1 indicates cluster location of similar levels. Oval 2 indicates cluster location of dissimilar disaster impact levels.}
\end{figure}

\subsubsection{\textit{Global spatial pattern}}
For computing spatial patterns and conducting the disparities analysis across 142 ZIP codes in Harris County, we included 12 business sectors (see Table 1), as their CCT variance curves present continuity and are  consistent match with the transitions in the resilience curve shown in Materials and Methods section. Using the disaster impacts and recovery duration for the 12 business sectors and their corresponding spatial lag values at each ZIP code, we illustrate the global Moran’s I scatterplot in Figs. S-41 and S-42 in the Supplementary Information. The values of global Moran’s I and their p-values are summarized in in Table 1. According to Table 1, we can see that the global Moran’s I related the disaster impacts in eight business sectors are significant (p-value < 0.1). Global Moran’s I for disaster impacts in drugstore, home supply, recreation and restaurant are not significant. Compared with disaster impacts, global Moran’s I for recovery duration are significant (p-value < 0.1) for four business sectors: cloth, drugstore, internet and telecommunication, and restaurant. For the eight remaining business sectors, global Moran’s I values are not significant. This result indicates that most business sectors (66.67\%) are more likely to present clusters of similar disaster impacts levels (i.e., positive values of global Moran’s I), while only 33.33\% showed clusters of similar recovery duration levels. Accordingly, hot spots with severe disaster impacts are more likely to be detected than hot spots with longer recovery duration for the defined business sectors.

\begin{table}[]
\caption{Global Moran’s I for disaster impact and recovery duration}
\centering
\begin{tabular}{llllll}
\hline
\multicolumn{3}{l}{Disaster Impact}                                                                        & \multicolumn{3}{l}{Recovery Duration}                                                                      \\ \hline
Business Sector                                                               & Global Moran’s I & p-value & Business sector                                                               & Global Moran’s I & p-value \\ \hline
Auto                                                                          & 0.19             & 0.002   & Auto                                                                          & 0.05             & 0.126   \\
Cloth                                                                         & 0.08             & 0.063   & Cloth                                                                         & 0.09             & 0.039   \\
Drugstore                                                                     & 0.04             & 0.196   & Drugstore                                                                     & 0.07             & 0.066   \\
Grocery                                                                       & 0.12             & 0.009   & Grocery                                                                       & 0.02             & 0.332   \\
Home supply                                                                   & -0.05            & 0.218   & Home supply                                                                   & -0.04            & 0.273   \\
\begin{tabular}[c]{@{}l@{}}Internet \&\\    \\ Telecommunication\end{tabular} & 0.08             & 0.058   & \begin{tabular}[c]{@{}l@{}}Internet \&\\    \\ Telecommunication\end{tabular} & 0.07             & 0.063   \\
Market                                                                        & 0.06             & 0.086   & Market                                                                        & -0.00            & 0.424   \\
Recreation                                                                    & 0.01             & 0.339   & Recreation                                                                    & -0.01            & 0.484   \\
Restaurant                                                                    & -0.02            & 0.413   & Restaurant                                                                    & 0.14             & 0.006   \\
Retail                                                                        & 0.08             & 0.067   & Retail                                                                        & -0.06            & 0.133   \\
Service                                                                       & 0.10             & 0.025   & Service                                                                       & 0.05             & 0.153   \\
Transportation                                                                & 0.07             & 0.082   & Transportation                                                                & -0.06            & 0.127   \\ \hline
\end{tabular}
\end{table}

Specifically, we can see the disaster impacts for two business sectors—home supply and restaurant—show negative values of global Moran’s I. Their global Moran’s I values, however, are not significant, meaning these two sectors show no spatial cluster potential for the disaster impacts. For the other 10 business sectors, disaster impacts show positive values of global Moran’s I. The global Moran’s I values of the disaster impacts for drugstore and recreation sectors are not significant as their p-values are larger than 0.1. Positive values of global Moran’s I mean that similar values are likely to be geographically clustered with each other. In our case, positive values of global Moran’s I for disaster impacts of the remaining eight business sectors reveal that the overall trend is that neighboring ZIP codes are more likely to have similar levels of disaster impact from Hurricane Harvey. 

In terms of recovery duration in Table 1, five business sectors presenting negative global Moran’s I values (home supply, market, recreation, retail and transportation) are not at significant levels (p-value > 0.1). For the seven remaining business sectors, the global Moran’s I values for only four of them are significant and positive. As a result, there is no significant spatial cluster trend for the recovery duration of most of business sectors: auto, grocery, home supply, market, recreation, retail, service and transportation. 

\subsubsection{\textit{Local spatial patterns}}
As for the global Moran’s I with Eq. (7), we computed the local Moran’s I with Eq. (8) for disaster impacts and recovery duration for the 12 business sectors. Using the grocery sector as an example, we present the geographic distributions of disaster impacts and recovery duration across the 142 ZIP codes in Harris County in Figs. 3a and 4a. Considering the statistical significance values, we also distinguished ZIP codes with local Moran’s I at high significance level (p-value < 0.1) in Figs. 3b and 4b. Local Moran’s I is significant in the black-shaded ZIP codes areas; areas shaded grey have local Moran’s I of low significance levels. To identify the locations of clusters with similar and dissimilar values at high significance level (p-value < 0.1), we plotted the ZIP codes with clusters of similar values, including high-high and low-low for disaster impacts and recovery duration, and those with dissimilar values (high-low and low-high) in Figs. 3c and 4c. The ZIP codes shaded blue and red are clusters of similar values; those shaded pink and light blue are clusters with dissimilar values.

\begin{figure}[hbt!]
\centering
\includegraphics[width=\linewidth]{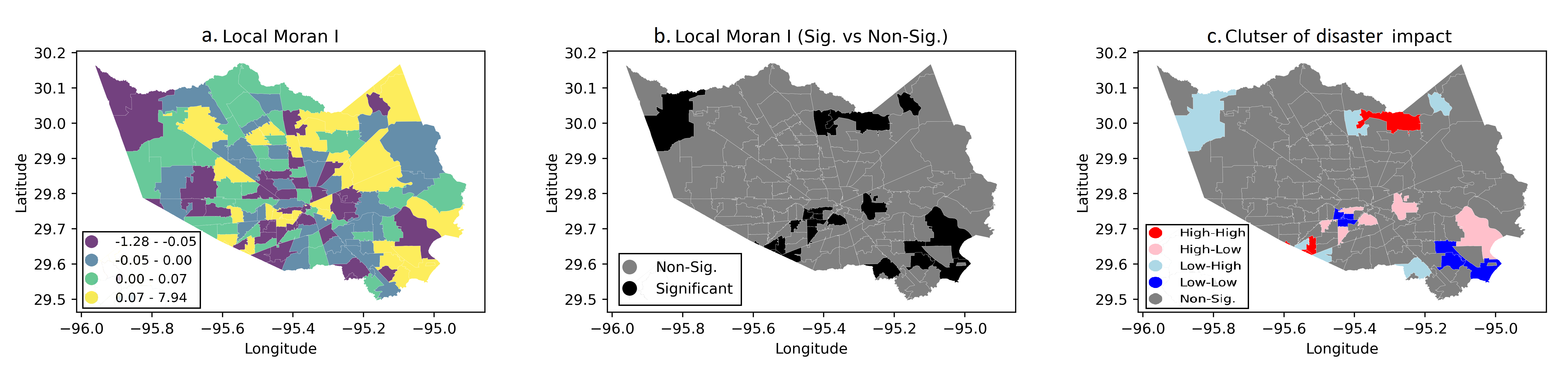}
\caption{LISA cluster map for disaster impact of grocery sector}
\end{figure}

\begin{figure}[hbt!]
\centering
\includegraphics[width=\linewidth]{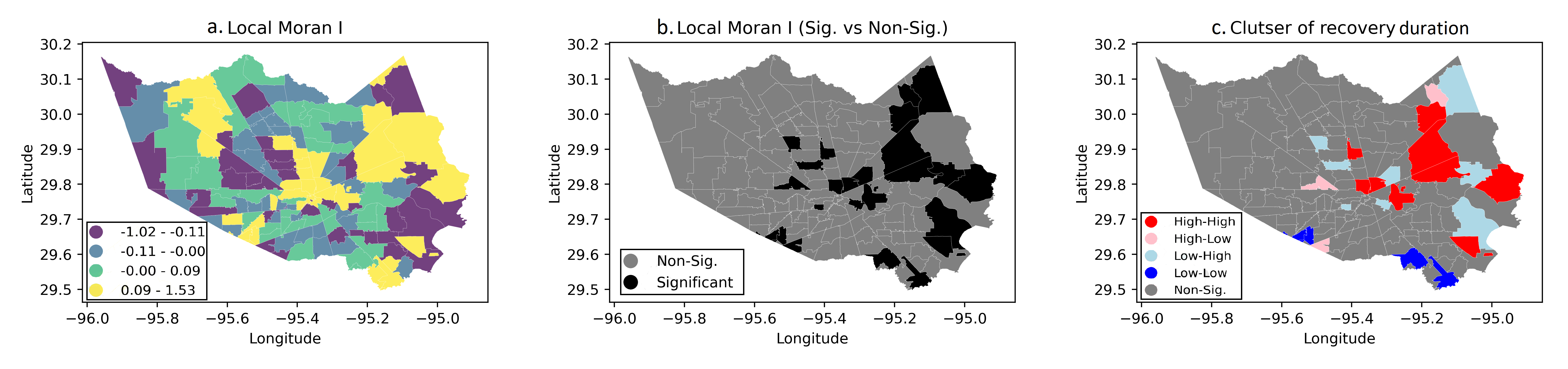}
\caption{LISA cluster map for recovery duration of grocery sector}
\end{figure}

For the 11 remaining business sectors, we present the Local Indicators of Spatial Association (LISA) maps in Figs. S19-40 of the Supplementary Information. Tables 2 and 3 summarize counts of clusters of ZIP codes of similar and dissimilar values with high significance levels (p-value < 0.1) for disaster impact and recovery duration for the 12 business sectors. For instance, for the grocery sector, among the 23 ZIP codes with local Moran’s I at significant level (p-value < 0.1) for disaster impacts, 11 ZIP codes are clusters of similarity. Seven of these 11 ZIP code regions are clusters of low disaster impact values (blue-shaded areas, Fig. 3), and four are clusters with high disaster impact values (red-shaded areas, Fig. 3). Since we used CCT variances based on changes in total transactions to quantify disaster impacts, their values are negative (indicating decline in transactions due to the disaster). Low values of disaster impacts refer to severe hurricane impact; high values represent slight hurricane impact (Fig. 2a). From the clusters of similarly low (or high) values for disaster impacts in the grocery sector, we can see that ZIP codes with severe (or slight) hurricane impacts are more likely to have neighboring ZIP code regions having suffered similar hurricane impacts. As the similarity cluster of low values accounts for 63.64\%, we can say the dominant similarity cluster type of disaster impacts in grocery sector is high-high. Hence, we can conclude that the dominant local correlation trend in disaster impacts of grocery sector is that ZIP codes with severe disaster impacts are more likely physically close to others with severe disaster impacts. Considering flood propagation characteristics, this trend can be expected as flooded regions are more likely to be located close to other flooded regions.

\begin{table}[]
\caption{Summary of LISA clusters count for disaster impact}
\centering
\begin{tabular}{lllllll}
\hline
Disaster impact                  & High-High & Low-High & Low-Low & High-Low & Sig. & Non-Sig. \\ \hline
Auto                             & 11        & 3        & 17      & 4        & 35   & 107      \\
Cloth                            & 5         & 8        & 25      & 6        & 44   & 98       \\
Drugstore                        & 9         & 3        & 16      & 11       & 30   & 103      \\
Grocery                          & 4         & 6        & 7       & 6        & 23   & 119      \\
Home supply                      & 0         & 3        & 7       & 11       & 21   & 121      \\
Internet    \& Telecommunication & 6         & 5        & 9       & 7        & 27   & 115      \\
Market                           & 6         & 3        & 17      & 12       & 38   & 104      \\
Recreation                       & 9         & 2        & 11      & 10       & 32   & 110      \\
Restaurant                       & 1         & 5        & 12      & 6        & 24   & 118      \\
Retail                           & 2         & 8        & 13      & 7        & 30   & 112      \\
Service                          & 12        & 2        & 6       & 9        & 29   & 113      \\
Transportation                   & 6         & 4        & 12      & 7        & 29   & 113      \\ \hline
\end{tabular}
\end{table}

\begin{table}[]
\caption{Summary of LISA clusters count for recovery duration}
\centering
\begin{tabular}{lllllll}
\hline
Recovery duration             & High-High & Low-High & Low-Low & High-Low & Sig. & Non-Sig. \\ \hline
Auto                          & 10        & 10       & 9       & 3        & 32   & 110      \\
Cloth                         & 12        & 12       & 7       & 2        & 33   & 109      \\
Drugstore                     & 9         & 11       & 4       & 2        & 26   & 116      \\
Grocery                       & 12        & 9        & 5       & 3        & 29   & 113      \\
Home supply                   & 0         & 11       & 7       & 3        & 21   & 121      \\
Internet \& Telecommunication & 10        & 12       & 4       & 3        & 29   & 113      \\
Market                        & 6         & 10       & 4       & 3        & 23   & 119      \\
Recreation                    & 4         & 6        & 6       & 2        & 18   & 124      \\
Restaurant                    & 6         & 10       & 6       & 1        & 23   & 119      \\
Retail                        & 2         & 12       & 3       & 4        & 21   & 121      \\
Service                       & 6         & 10       & 7       & 1        & 24   & 118      \\
Transportation                & 4         & 11       & 4       & 4        & 23   & 119      \\ \hline
\end{tabular}
\end{table}

For LISA clusters of disaster impacts for other business sectors, we can use the same analysis process to derive their cluster characters. According to Table 2, the dominant cluster patterns for the disaster impacts of various business sectors are as follows: 1) auto: low-low (ZIP codes with severe hurricane impacts are more likely to have neighboring ZIP codes with severe hurricane impacts); 2) cloth: low-low; 3) drugstore: low-low; 4) home supply: high-low (ZIP codes with slight hurricane impacts are more likely to have neighboring ZIP codes with severe hurricane impacts); 5) internet and telecommunication: low-low; 6) market: low-low; 7) recreation: low-low; 8) restaurant: low-low; 9) retail: low-low; 10) service: high-high (ZIP codes with slight hurricane impacts are more likely to have neighboring ZIP codes with slight hurricane impacts); 11) transportation: low-low. As a result, the dominant LISA cluster pattern is low-low for the disaster impacts for 10 out of 12 business sectors.

In terms of the LISA clusters of recovery duration, Table 3 illustrates the dominant cluster patterns: 1) auto: none; 2) cloth: none; 3) drugstore: low-high (ZIP codes with short recovery duration are more likely to have neighboring ZIP codes with long recovery duration); 4) grocery: high-high (ZIP codes with long recovery duration are more likely to have neighboring ZIP codes with long recovery duration); 5) home supply: low-high; 6) internet and telecommunication: low-high; 7) market: low-high; 8) recreation: none; 9) restaurant: low-high; 10) retail: low-high; 11) service: low-high; 12) transportation: low-high. Therefore, the dominant LISA cluster pattern is low-high for the recovery duration of 8 out of 12 business sectors. The results indicate that ZIP codes with sufficient recovery resources (i.e., short recovery duration) are surrounded by ZIP codes with insufficient recovery resources (i.e., long recovery duration). This result indicates the spatial heterogeneity of short-term recovery duration.

Tables 2 and 3 illustrate that LISA cluster counts for disaster impact range from 21 to 44, and for recovery duration, 18 to 33, at significant level (p-value < 0.1). Based on the flood map from Federal Emergency Management Administration (FEMA), we identified the 35 ZIP codes, clustered in southwest Harris County, where flood claims were reported (red-shaded areas, Fig. 5). ZIP codes in southwest of Harris County present significant LISA clusters for the disaster impacts of grocery sector, which is consistent with the flood situations in Fig. 5. Conversely, we expect that disaster impacts and recovery duration of ZIP codes outside the flooded areas do not show significant LISA clusters. 

\begin{figure}[ht]
\centering
\includegraphics[width=0.6\linewidth]{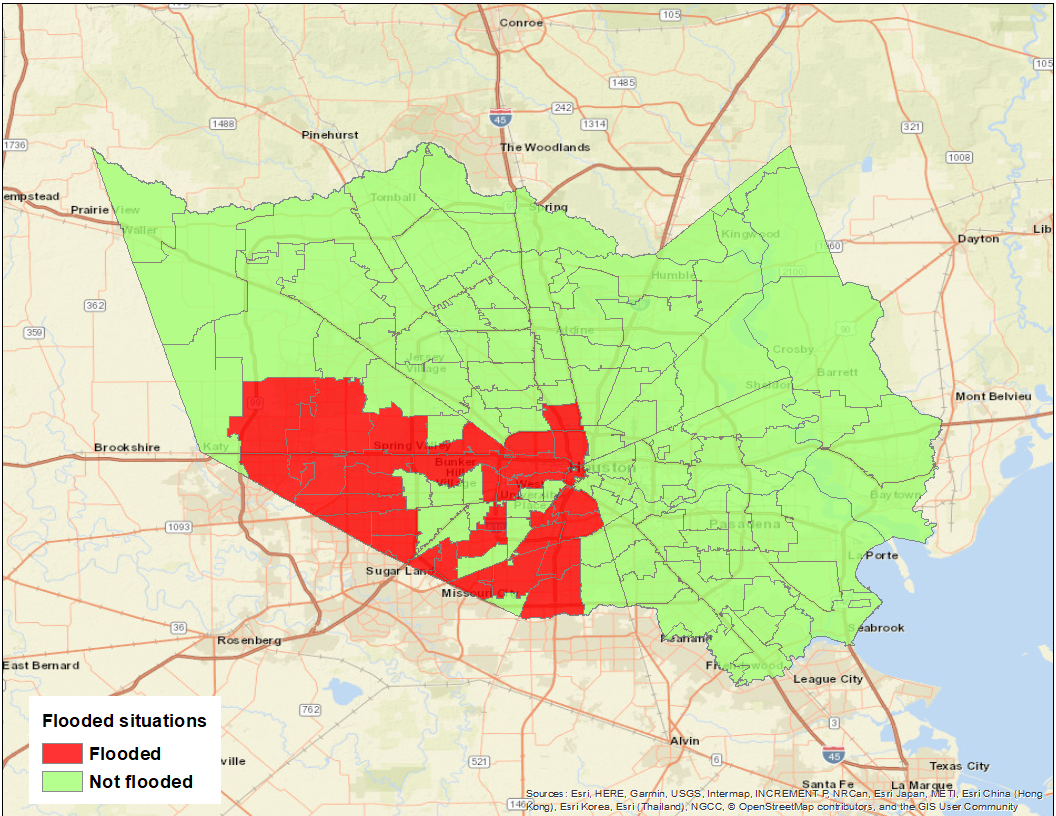}
\caption{Flood map for ZIP codes in Harris County}
\end{figure}

\subsection{Spatial disparities analysis results}
\subsubsection{\textit{Statistics of disaster impact and recovery duration}}
In this section, we first describe the spatial disparities of disaster impact and recovery duration for the defined business sectors across the 142 ZIP codes in Harris County. The summary of statistical analysis—mean, median, and standard deviation—of disaster impact and recovery duration is presented in Table 4. According to the results related to disaster impact in Table 4, the home supply sector has the largest mean, median, and standard deviation (std = 0.53), while the restaurant sector has the lowest standard deviation value (std = 0.12). This result means that the home supply sector has seen the most significant spatial variations among the ZIP codes in terms of disaster impact, and the restaurant sector has shown the least significant spatial variations. For recovery duration, the internet and telecommunication sector shows the largest mean and median values. In addition, we can see cloth sector possesses the greatest standard deviation (std = 6.82), while recovery duration for the grocery sector is the lowest (std = 3.49). This result reveals that the recovery duration of the cloth sector varies most significantly across ZIP codes, while the recovery duration of grocery sector varies the least across the ZIP codes among the 12 business sectors in Harris County.

% \usepackage{multirow}
\begin{table}[]
\caption{Summary of spatial disparities among disaster impact and recovery duration by ZIP code regions}
\centering
\begin{tabular}{lllllll}
\hline
\multirow{2}{*}{Business sector}                                              & \multicolumn{3}{l}{Disaster impact} & \multicolumn{3}{l}{Recovery duration (days)} \\ \cline{2-7} 
                                                                              & mean       & median      & std      & mean          & median         & std         \\ \hline
Auto                                                                          & -0.18      & -0.14       & 0.18     & 4.82          & 3.5            & 4.76        \\
Cloth                                                                         & -0.80      & -0.81       & 0.26     & 10.35         & 9              & 6.82        \\
Drugstore                                                                     & -0.45      & -0.41       & 0.27     & 6.09          & 5              & 4.64        \\
Grocery                                                                       & -0.37      & -0.33       & 0.22     & 5.62          & 5              & 3.49        \\
Home   supply                                                                 & -0.96      & -1.04       & 0.53     & 6.09          & 4.5            & 6.08        \\
\begin{tabular}[c]{@{}l@{}}Internet \&\\    \\ Telecommunication\end{tabular} & -0.45      & -0.41       & 0.23     & 11.20         & 10             & 6.23        \\
Marketing                                                                     & -0.18      & -0.14       & 0.18     & 4.82          & 3.5            & 4.76        \\
Recreation                                                                    & -0.57      & -0.54       & 0.36     & 5.40          & 4              & 5.97        \\
Restaurant                                                                    & -0.46      & -0.46       & 0.12     & 7.84          & 7              & 3.58        \\
Retail                                                                        & -0.55      & -0.51       & 0.30     & 6.78          & 5              & 5.65        \\
Service                                                                       & -0.46      & -0.39       & 0.29     & 6.06          & 5              & 5.22        \\
Transportation                                                                & -0.78      & -0.73       & 0.40     & 7.81          & 6.5            & 6.12        \\ \hline
\end{tabular}
\end{table}

\subsubsection{\textit{Spatial disparities}}
For spatial disparities of disaster impact and recovery duration, we conducted a stepwise regression model selection. The modeling approach was implemented in our regression analysis for disaster impact (absolute value) and recovery duration of the 12 business sectors. In this section, the stepwise regression results for disaster impact and recovery duration of drugstores are presented in Tables 5 and 6, respectively. The regression analysis results for disaster impact and recovery duration of other business sectors are presented in Tables S2–23 in the Supplementary Information.

\begin{table}[]
\caption{Gaussian regression analysis for disaster impact on the drugstore sector}
\centering
\begin{tabular}{lllll}
\hline
Variables        & Coefficient & Std. Error & P-values         & VIF  \\ \hline
Intercept        & 0.89        & 0.24       & \textless{}0.001 & -    \\
Reported loss    & 2.63 e-05   & 1.87 e-05  & 0.16             & 1.13 \\
Total population & -3.32 e-06  & 1.16 e-06  & 0.004            & 1.04 \\
Age              & -1.51 e-02  & 7.71 e-0-  & 0.05             & 1.84 \\
Income           & 5.09 e-06   & 1.35 e-06  & \textless{}0.001 & 1.67 \\ \hline
\end{tabular}
\end{table}

\begin{table}[]
\caption{Gaussian regression analysis for recovery duration on the drugstore sector}
\centering
\begin{tabular}{lllll}
\hline
Variables        & Coefficient & Std. Error & P-values         & VIF  \\ \hline
Intercept        & 1.82        & 0.12       & \textless{}0.001 & -    \\
Number of claims & 9.03        & 2.21       & \textless{}0.001 & 1.12 \\
White (\%)       & 0.28        & 0.19       & 0.136            & 1.10 \\
Income           & -5.45 e-06  & 1.83 e-06  & 0.003            & 1.21 \\ \hline
\end{tabular}
\end{table}

Regression results for disaster impact of the drugstores sector (Table 5) include reported loss of the residents, total population of the ZIP code, median age, and income level of the residents. The model results suggest that a larger amount of reported loss is positively related to the disaster impact (based on change in transactions in drugstores). This result means that ZIP codes with larger amounts of total loss sustained a greater disaster impact for the drug store business sector. The reported loss is an indicator of the extent of direct physical damage in the affected area. Since most drugstores are visited by local residents, this result could be due to drugstores being flooded, residents losing access to drugstores, or residents relocating to temporary housing due flooding of residences. Total population and median age within the ZIP codes are negatively associated with the disaster impact. Those ZIP codes with a higher population and areas with a higher median age experienced a lesser disaster impact related to drugstores. The results show that areas with a higher population experienced less impact during Hurricane Harvey. With a higher population, more stores may exist within the ZIP code area, providing redundancy and lower impact. Furthermore, ZIP Code areas with a higher percentage of the elderly experienced less disaster impact. Elderly populations may have a higher need for the drugstores and make more drugstore-related purchases from drugstores in other areas; therefore, the greater transactions for this category could contribute to the less disaster impact upon the drugstore sector. Finally, areas with higher income experienced a higher level of impact for the transactions related to drugstores, as suggested by the positive relationship between income and the disaster impact. This result implies that wealthier residents could continue fulfilling their drugstore-related needs even if they live in a flood-affected area.

Table 6 shows the Poisson regression results for the recovery duration related to drugstores transaction. Income has a significant and negative relationship with the recovery duration of the drugstore sector. Higher-income residents recovered quicker as indicated by their drugstore-related transactions. Considering the significant and positive relationship between the income and the disaster impact, we can see that although the communities with higher incomes experienced a greater impact, they recovered more quickly than others. Such a pattern is also observed in other business sectors, which highlights the influences of socioeconomic factors in the resilience of residents in response to disaster-induced perturbations. The results also suggest that those areas with a larger number of flood insurance claims from the disaster required greater recovery duration. Similar to the relationship between loss and the disaster impact, this association could be explained by the higher physical damage in areas with a higher reported claim in comparison with other regions. The percentage of the white residents has a positive relationship with recovery duration level. The results indicate that ZIP codes with a higher white population percentage took longer to recover based on drugstore transaction changes. ZIP codes with a higher percentage of white population (such as neighborhoods near the reservoirs) were the most seriously affected by flooding during Hurricane Harvey. The more severe physical damage to the drugstore sector caused by flooding could explain the slower recovery in ZIP codes with large white populations. The VIF (variance inflation factor) values for all variables included in the model are less than 2 which suggest that the models are not affected by the multicollinearity issue. 

\subsection{Disaster impact and recovery duration for various business sectors }
\subsubsection{\textit{Drugstore and health care}}
The drugstore sector refers mainly to drugstores and pharmacies, while the health care sector represents hospitals, medical service centers, and doctors and physicians. Both sectors provide community residents with medical resources to cope with disasters. CCT variances based on the total transaction values for drugstores are illustrated in (Fig. 6a) and health care (Fig. 6b). Fig. 6a documents a significant increase in the total transactions in the drugstore sector before Hurricane Harvey made landfall in Harris County (red vertical line). Immediately after landfall, a sharp decrease occurs. This phenomenon may reflect the possibility that community residents in Harris County used their credit cards to purchase necessary medicines from the drugstores for preparing the response to Hurricane Harvey. We can see the same situation in the CCT variance curve of the grocery sector (Fig. S-4 in Supplementary Information) as community residents stored foods and other supplies before the hurricane made landfall. Transactions for the purchase of medications show an immediate drop after the hurricane made landfall; recovery took about one and a half weeks to the new steady level. Also, the new steady level is higher than the previous baseline level. This phenomenon indicates that community residents made more transactions in the drugstore sector in the recovery period, starting September 9, 2017. Compared with the drugstore sector, no discernable increase of health care transactions occurred before Hurricane Harvey made landfall in Harris County (Fig. 6b). This situation is reasonable as medical services from either hospitals or medical centers cannot be stockpiled to prepare for a hurricane. Fig. 6b indicates sharp decreases in total transactions after landfall of the hurricane, with overall recovery to baseline level occurring over the course of two weeks.

In addition, the disruption (orange line segment) and recovery (green line segment) periods of drugstore and health care sectors are marked in Figs. 6a and 6b. Using the maximum drop of total transactions from their baseline levels illustrated in their 7-day moving average curves (Eq. (3)), we quantify the disaster impacts on drugstore and health care sectors. We added the periods of the orange and green line segments to calculate the recovery duration (Eq. (4)) for total transactions to return to their baseline levels.

\begin{figure}[ht]
\centering
\includegraphics[width=\linewidth]{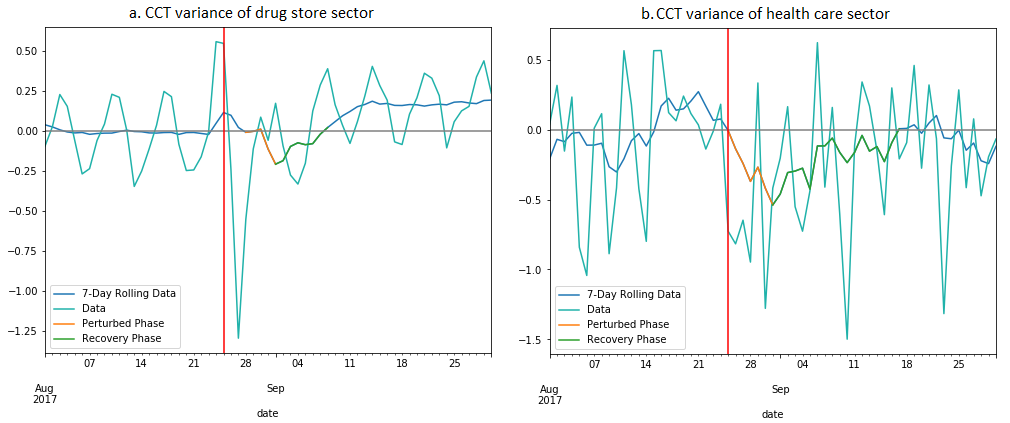}
\caption{CCT variance curves for drug store (6a) and health care (6b) sectors. In both figures, the cyan curve represents the daily CCT variance from rough data (Eq. (1)), while the blue curve shows the CCTV by the 7-day moving average (Eq. (2)). Specifically, the golden part of the blue curve reveals the perturbed stage (i.e., disruption stage) while the green part reflects the recovery stage. The red vertical line marks the date when Hurricane Harvey made landfall in Harris County. The grey horizontal line highlights where the total transactions equal to their corresponding baseline values. These descriptions apply to Fig. 7 and Figs. S1-18 in the Supplementary Information.}
\end{figure}

\subsubsection{\textit{Utilities–electric, gas, water, and sanitary}}
The utilities sector mainly includes the supply of electricity, gas, water, and sanitary, which plays a critical role to support community residents’ daily life needs in both normal and disaster periods (39–41). Fig. 7 shows the CCT variance patterns by community residents’ transactions in the utilities sector. There is not a significant variation from the normal to hurricane periods. Hence, this result cannot quantify the disaster impacts and recovery duration based on the total transactions in the utilities sector. This result could be explained by utility customers continuing to pay for their utilities as long as their homes were not flooded.

\begin{figure}[ht]
\centering
\includegraphics[width=0.6\linewidth]{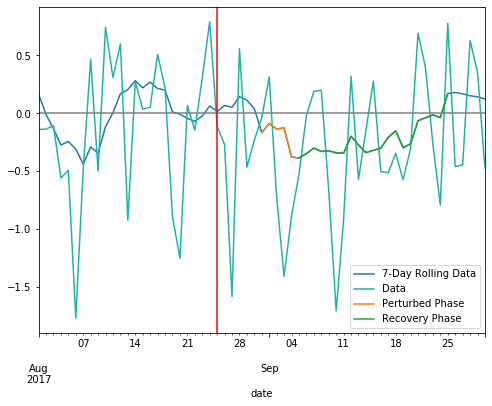}
\caption{CCT variance curve for utilities sector.}
\end{figure}

\subsubsection{\textit{Summary of disaster impacts and recovery duration}}
Using the CCT curves for all defined business sectors (e.g., Figs. 6-7), we computed their disaster impacts and recovery duration. The results of disaster impacts (left) and recovery duration (right) for these sectors are presented in Fig. 8. As the CCT variance curves for 7 of the 21 business sectors are not continuous and are not following the resilience transition patterns shown in Fig. 10, we quantified the disaster impacts and recovery duration for the 14 remaining business sectors. The names of these 14 business sectors are illustrated in Fig. 8.

\begin{figure}[ht]
\centering
\includegraphics[width=\linewidth]{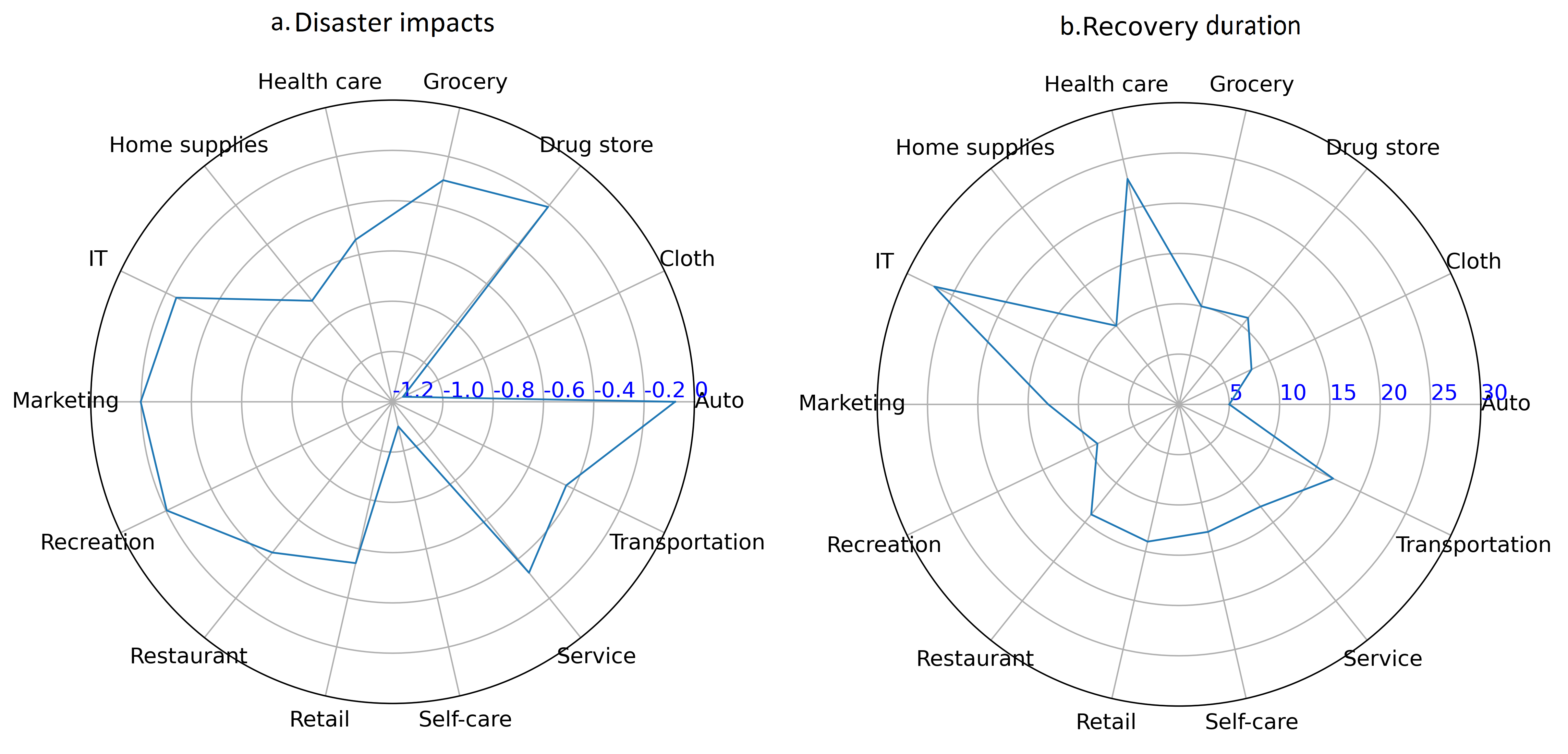}
\caption{Disaster impact (8a) and recovery duration (8b) for the defined business sectors at county level.}
\end{figure}

In Fig. 8a, we can see business sectors including cloth, home supplies, and self-care had the most severe disaster impacts. This situation can be expected as purchase of clothes and home supplies (e.g., furniture, home furnishing, and household appliances) and consumption in self-care services (e.g., dental and optician services) are not daily life needs. For the recovery duration in Fig. 8b, the recovery of the internet telecommunication sector has taken the longest time. This is consistent with a survey result by Podesta et al.(42), which shows that the recovery time of internet ranged from a few days to four weeks after Hurricane Harvey in Harris County. In addition, the recovery durations for auto, cloth, and recreation sectors show short periods. Fig. 8a illustrates slight disaster impacts on auto and recreation sectors, which can explain the fast recovery of these sectors. An interesting result is related to the internet and telecommunication sector; by juxtaposing both figures in Fig. 8, we can see that this sector had slight disaster impacts but long recovery duration. This situation can show that disaster impacts and recovery duration are not always positively correlated across all business sectors.

\section{Discussions and Conclusion}
Credit card transaction (CCT) variance data can reveal the collective impacts of disasters on community residents, business, and infrastructure systems. Disaster impacts on community residents and their households can influence their consumption in business sectors, such as cloth, retail, and home supplies. Meanwhile, business closures due to disaster impacts, such as closures of restaurants and recreation services, can also reduce or limit community residents’ consumptions in these business services. Also, disaster impacts on infrastructure systems, such as road closures due to floods, can reduce residents’ access to business services and further affect consumptions. Therefore, we examined the changes in the state of community in terms of disaster impacts and recovery duration with CCT variance data from both community residents’ perspective at ZIP code scale and businesses’ perspective at county scale. At the ZIP code scale, we also inspected the spatial patterns and disparities of these community state indices. The results show the value of using human activity data (such as credit card transactions) for capturing the combined effects of household impacts, infrastructure disruptions, and business impacts as part of community resilience evaluations.

From community residents’ perspective at the ZIP code scale, we utilized CCT variance data to examine the spatial patterns of disaster impacts and recovery duration of communities across the ZIP code regions of Harris County using global and local spatial autocorrelation tools. According to analysis results from these tools, we find that the overall pattern of disaster impacts for the auto, cloth, grocery, internet and telecommunication, market, retail, service, and transportation sectors is a cluster of similar disaster impact levels. With LISA cluster maps, our results identify clusters with severe disaster impacts or long recovery duration. For instance, the blue-shaded areas representing ZIP code areas in Fig. 3c refer to the locations of clusters with severe disaster impact of the grocery sector, while red-shaded areas in Fig. 4c mark locations of clusters with long recovery duration. Our findings in spatial patterns can benefit emergency management agencies to identify hot spots with severe disaster impacts and long recovery duration for specific business sectors and further support their design and implementation of recovery strategies. 
In addition, this research explores factors which may influence the spatial disparities of disaster impacts and recovery duration of communities in Harris County. We find that factors of reported loss, total population, age, and household income may explain the spatial disparities of disaster impacts for the drugstore sector. Meanwhile, variables such as number of insurance claims, percentage of the white population, and household income can explain the spatial disparities of the recovery duration for drugstore sector. Our results in spatial disparities analysis can benefit urban planners to identify dominant factors influencing disaster impacts and recovery duration of specific business sectors and further build the long-term resilience for our cities. 
From the business perspective at the county scale, we evaluated the disaster impacts and recovery duration of business sectors. We found that business sectors providing resources for community residents’ daily life needs, such as drugstores, experienced slight disaster impact levels. Also, business sectors with slight disaster impacts, auto and recreation, showed short recovery duration. For some sectors such as the internet and telecommunication sector, however, the recovery duration was long despite the slight disaster impacts level. These results indicate that disaster impacts and recovery duration are not always positively correlated across various business sectors.

A limitation in this study falls in the lack of proper explanations for the observed CCT variances across the hurricane period, as credit card transactions could be impacted by various additional factors, such as major community events or holidays (e.g., Labor Day on September 4, 2017). Future research will investigate the local news and holiday/festival calendars across communities and make the corresponding adjustments on the CCT to normalize the impacts from the hurricane on the state of community. Also, future research can explore the CCT in the same period of 2016 to use as a baseline to compute the CCT variances.

Despite limitations, this research contributes to the body of knowledge related to community resilience evaluation through two main aspects: first, the study and results demonstrate the potential for utilizing human activity data for more integrative evaluation of the spatial and temporal patterns of impacts and recovery in community resilience assessments. Second, the approach could be utilized for enhancing situational awareness to better inform resource allocation and recovery strategy prioritization efforts by emergency managers and public agencies.

\section{Materials and Methods}

The methodological steps for analyzing the disaster impacts and short-term recovery in this study are illustrated in Fig. 9. In this study, we quantified two indices of community state—disaster impact and recovery effort—as suggested by Vugrin et al. (1) across ZIP codes in Harris County. Since this research relied on the number of days to reveal the recovery effort, we employed recovery duration to represent the recovery effort. We also conducted spatial autocorrelation analyses to reveal spatial patterns of disaster impacts and recovery duration of communities. Subsequently, we utilized the stepwise regression model to examine spatial disparities in disaster impacts and recovery duration with respect to demographic and flood claim data across different ZIP codes regions. Finally, we quantified the disaster impacts and recovery duration of various business sectors at the county level in Harris County. 

\begin{figure}[ht]
\centering
\includegraphics[width=0.8\linewidth]{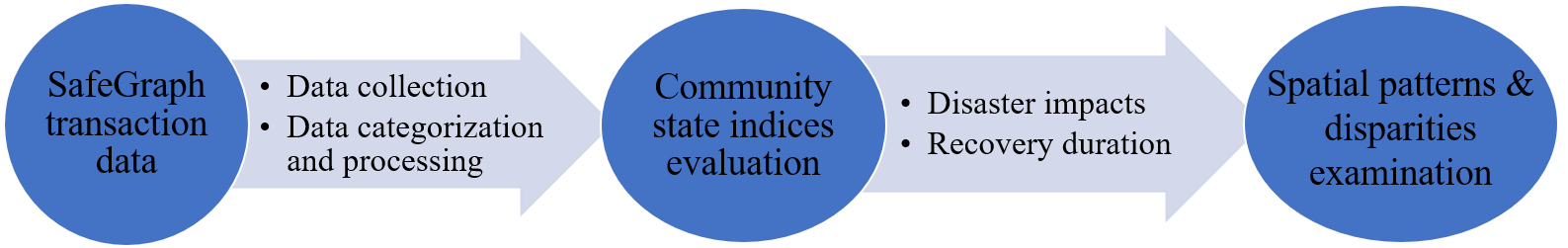}
\caption{Research framework for examining disaster impacts and recovery duration of community.}
\end{figure}

\subsection{SafeGraph transaction data}
\subsubsection{\textit{Data description}}
The credit card transaction (CCT) data was provided by Safegraph company. Each item within the CCT data contains the transaction date, cardholder’s residential ZIP code, the number of unique cards from the ZIP code involved in transactions on that day, the number of unique transactions on cards from the ZIP code observed on that day, and the total amount spent on cards from the ZIP code on that day (Table 7).

\begin{table}[h]
\caption{Descriptions of the CCT data and implementations in this research}
\centering
\begin{tabular}{lll}
\hline
Attribute                                                           & Contents                                                                                                                         & Use in this research                                                                                                         \\ \hline
Date                                                                & Year-month-day represents   transaction day.                                                                                     & Filter the CCT   data with defined timeline.                                                                                 \\
Cardholder Zip                                                      & The cardholder’s   residential ZIP code.                                                                                         & Filter the CCT   data in Harris County.                                                                                      \\
Number of cards                                                     & \begin{tabular}[c]{@{}l@{}}The number of   unique cards from that ZIP code that \\ transacted on that day.\end{tabular}          & \begin{tabular}[c]{@{}l@{}}Reflect variances   in CCT active users (not \\ used for resilience analysis).\end{tabular}       \\
\begin{tabular}[c]{@{}l@{}}Number of   \\ transactions\end{tabular} & \begin{tabular}[c]{@{}l@{}}The number of   unique transactions on cards from \\ that ZIP code observed on that day.\end{tabular} & \begin{tabular}[c]{@{}l@{}}Reveal variances   in CCT in defined \\ timeline (not used for resilience analysis).\end{tabular} \\
Total   spent                                                       & \begin{tabular}[c]{@{}l@{}}The   total amount spent on cards from this ZIP code \\ on that day.\end{tabular}                     & Reveal variances in CCT in defined timeline.                                                                                 \\ \hline
\end{tabular}
\end{table}

This study encompassed the period from August 1, 2017, through September 30, 2017. Hurricane Harvey made landfall in Harris County on August 24, 2017. The analysis period includes the steady-state period (prior to Harvey landfall), perturbed period (from landfall until floodwater receded), and short-term recovery period. Based on the ZIP code list in Harris County (35), we filtered the CCT data within the study region. The CCT data related to 142 ZIP codes in Harris County was aggregated and further used for quantifying disaster impacts and recovery duration across ZIP codes. In addition, the total amount spent on cards was used in calculating variances in the CCT data. The descriptions of the CCT data and their use in different steps of this research are presented in Table 7. Since this research focused on the aggregated-level CCT variances in Harris County, we excluded the attribute “number of cards” in Table 7 from the analysis. 

\subsubsection{\textit{Data categorization and process}}
The original CCT data contained 396 MCCs (merchant category codes). Each MCC consists of three to four digits representing a business category as defined by the Visa Merchant Data Standards Manual (36). We categorized these 399 MCCs into 21 business sectors:1) auto, 2) cloth, 3) drugstore, 4) financial, 5) governmental, 6) grocery, 7) health care, 8) home supply, 9) hotel, inns, and lodging, 10) insurance, 11) internet and telecommunication, 12) marketing, 13) organization, 14) recreation, 15) restaurant, 16) retail, 17) school, 18) self-care, 19) service, 20) transportation, and 21) utilities–electric, gas, water, and sanitary. Each business sector includes several MCCs, as illustrated in the Supplementary Information (Table S-1). Based on the 21 defined business sectors, we aggregated the total spent by dates and sectors. The daily total spent was applied for revealing the CCT variances from the steady-state period (prior to hurricane periods) in Harris County.

\subsection{Community resilience evaluation indices}
Vugrin et al. (1) proposed a framework to quantify two components (i.e., disaster impact and recovery effort) in resilience assessments. Disaster impact is defined as the difference between a targeted system performance level and an actual system performance following a disruptive event, which can be measured by the variances in system performance from normal to disruption periods1. Recovery effort refers to the quantity of resources used in the recovery process after the disruption. In the context of this study, recovery effort is measured based on the number of days or transaction levels to return to pre-disaster levels. We used recovery duration as a proxy for overall community recovery. In this analysis, we considered the level of credit card transactions as a measure of the state of the community. Accordingly, based on changes in the transaction levels, we computed disaster impacts and recovery duration across ZIP codes and business sectors.

\subsubsection{\textit{Disaster impact}}
To illustrate the approach for computing disaster impacts using CCT variance data, we first explained community resilience transitions and phases (Fig. 10). As explained earlier in this study, we examined CCT data as a measure for the state of the community and as a metric to evaluate variations in transactions prior to, during, and after a disaster to determine transitions in the state of the community in four stages: initial steady stage, disruptive/perturbed stage, recovery stage, and new steady stage. Two recovery stages are shown in Fig. 10. The solid curve represents that the state of the community recovery to the pre-disruption steady-state level. The dashed curve represents the situation of the state of community recovering to a lower level than its initial steady stage. Accordingly, the extent of disaster impacts and recovery duration could be determined based on metrics proposed by Nan and Sansavini (37).

\begin{figure}[ht]
\centering
\includegraphics[width=0.65\linewidth]{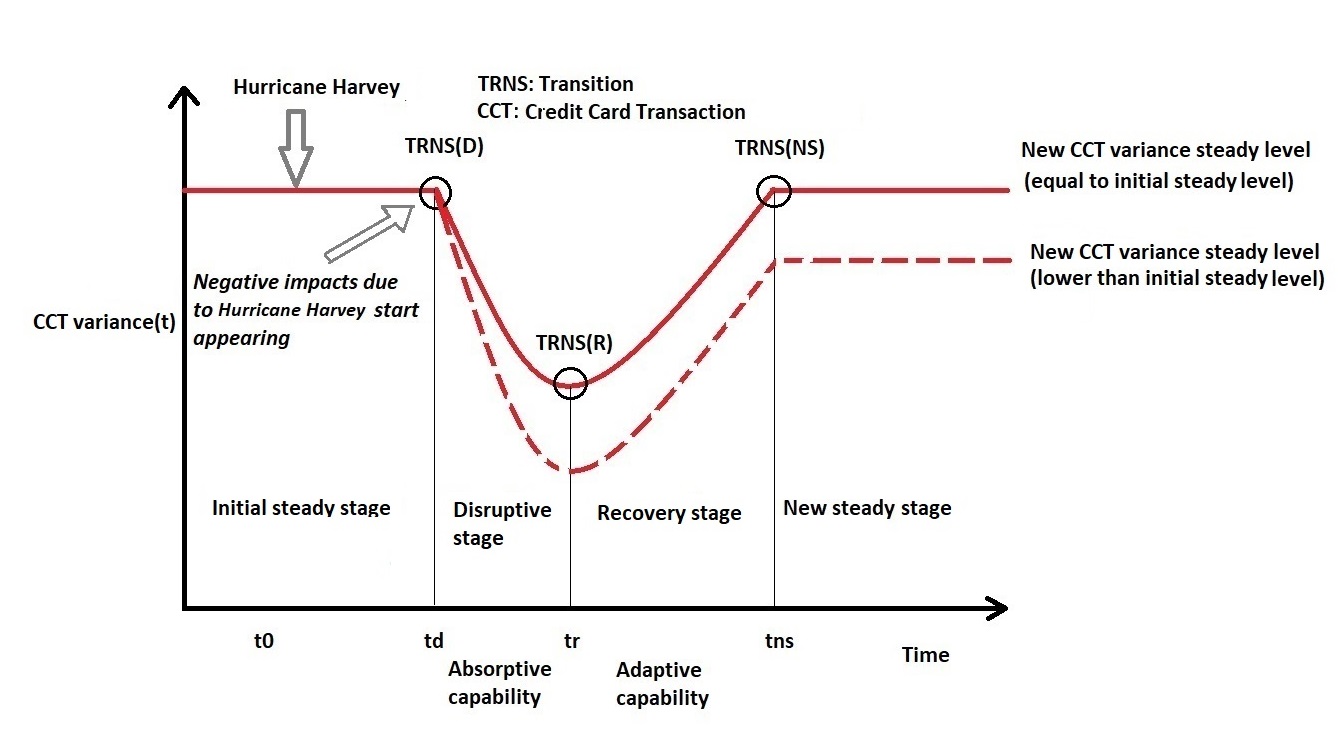}
\caption{Community resilience transitions and phases by CCT variance data (adapted from Nan and Sansavini (37)).}
\end{figure}

Specifically, the CCT variance (t) in Fig. 10 means the varying percentage of CCT at time t compared with the CCT baseline. We considered the CCT data related to the first three weeks of August to calculate the baseline, as the level of transactions was relatively steady for each business sector before Hurricane Harvey made landfall in Harris County. We calculated the baseline for each day of the week separately, and each business sector with its own seven baseline values (one baseline value for each day of the week). For example, the average of total transactions occurred on each Monday was calculated as the baseline value for total transactions on Monday. Similarly, we computed other baseline values for total transactions for other days of a week. Subsequently, we quantified the daily CCT variance in terms of total transactions using Eq. (1).
\begin{equation}
\label{eq:1}
TS_percentage\ variance(t)=\ \frac{total\ transaction(t)\ -\ ts_baseline\ value}{\frac{total\ transaction(t)\ +\ ts_baseline\ value}{2}}
\end{equation}
where, t represents the transaction date; ts\_baseline\ value: the baseline value of total transaction for the day t within a week.

In addition, to reduce the single event impact on people’s daily CCT behavior, such as Labor Day on September 4, 2017, we calculated the seven-day moving average to quantify the disaster impacts and recovery effort metrics for the 21 defined business sectors. The CCT variance based on the seven-day moving average is obtained using Eq. (2):
\begin{equation}
\label{eq:2}
TS7_percentage\ variance(t)=\ \frac{\ \frac{\sum_{t-6}^{t}{total\ transaction(i)}}{7}-\ ts_baseline\ value}{\frac{\frac{\sum_{t-6}^{t}{total\ transaction(i)}}{7}\ +\ ts_baseline\ value}{2}}
\end{equation}

Accordingly, we calculated the disaster impacts metric for the 21 business sectors with the seven-day tolling mean values of CCT variance using Eq. (3):
\begin{equation}
\label{eq:3}
disaster\ imapct_TS7=\ TS7_percentage\ variance(t_r)\ -\ TS7_percentage\ variance(t_d)
\end{equation}
where, $t_d$ represents the transaction date when negative impacts occurred due to Hurricane Harvey; $t_r$ represents the transaction date when the maximum negative impacts occurred.

\subsubsection{\textit{Recovery duration}}
Recovery effort can be quantified by the quantity of resources consumed or duration of the recovery process after the disaster disruptions. In this study, we used the recovery duration to reflect the recovery effort for the 21 defined business sectors. Using the variances related to the total transactions (i.e., CCT variance(t), Fig. 10), we calculated the recovery duration with Eq. (4).
\begin{equation}
\label{eq:4}
recovery\ duration=\ t_{ns}\ -\ t_d
\end{equation}
where, $t_d$ means the time when negative impact of disruptive events present; $t_{ns}$ represents the time when the system recovers to a new steady state.

\subsection{Spatial patterns of disaster impact and recovery duration}
In the next step, we employed spatial autocorrelation analysis to investigate the spatial patterns of disaster impacts and recovery duration across ZIP codes and business sectors. Our objective was to examine the spatial relationship among ZIP codes in terms of their disaster impacts and recovery duration. Spatial autocorrelation reveals the presence of spatial clustering (such as hot spots) for the value of the variable of interest (i.e., disaster impacts and recovery duration in this study) in a given location. Positive values mean similarity and geographical closeness are present simultaneously for the value of the variable of interest, while negative values indicate similar values are distributed far from each other in a region. 

Two commonly used approaches for spatial auto-correlation analysis are the global and local processes. The global spatial autocorrelation can detect the spatial clustering of disaster impacts and recovery duration. However, the global approach cannot specify the location of the clusters (such as the areas with clusters of high and low disaster impacts or long and short recovery duration). Hence, we also conducted the local spatial autocorrelation to identify the cluster locations for both high and low values of these community state indices. The steps for conducting the global and local spatial autocorrelation are explained below.

\subsubsection{\textit{Spatial lag}}
Prior to conducting the global and local autocorrelation analyses, we first calculated the spatial lag for each of the two indices (i.e., disaster impacts and recovery duration) for the 21 defined business sectors. Spatial lag refers to the variable that averages the neighboring values of a location (38). For instance, if region A has shared boundaries with regions B (recovery duration: 12 days), C (recovery duration: 15 days), D (recovery duration: 12 days) and E (recovery duration: 18 days), the spatial lag of recovery duration for region A will be the weighted average of 12 (region B), 15 (region C), 12 (region D), and 18 (region E). This calculation started with building the spatial weight matrix W. We used the queen contiguity weight, which reveals adjacency relationships with a binary indicator variable: 1 means a polygon shares an edge or a vertex with another polygon, while 0 means a polygon does not share edges or vertexes with another polygon. We used the row-standardized W, where values range from 0 to 1. For instance, if a ZIP code area polygon i shares an edge with four polygons, j, k, l and m, the values of w{ij}|, w{ik}|, w{il}| and w{im}| will be 0.25. With established row-standardized weight matrix W, we computed the spatial lag of disaster impacts and recovery duration for all the ZIP codes in Harris County using Eq. (5-6).
\begin{equation}
\label{eq:5}
{DI}_{sl-i}\ =\ \sum_{j}{w_{ij}{DI}_j}
\end{equation}

\begin{equation}
\label{eq:6}
{RD}_{sl-i}\ =\ \sum_{j}{w_{ij}{RD}_j}
\end{equation}
where $w_{ij}$ represents the spatial weight of ZIP code j on ZIP code i; ${DI}_j$: disaster impacts in ZIP code j; ${RD}_j$: recovery duration in ZIP code j; ${DI}_{sl-i}$: spatial lag of disaster impacts in ZIP code i; ${RD}_{sl-i}$: spatial lag of recovery duration in ZIP code i.

\subsubsection{\textit{Global spatial autocorrelation}}
We used Moran’s I to detect the clustering of disaster impacts and recovery duration. First, we used Moran Plot for the visualization of disaster impact and recovery duration and their corresponding spatial lag values for the defined business sectors. This plot can reveal the nature and extent of spatial autocorrelation for the two indices of disaster impacts and recovery duration. Then, we computed the global Moran’s I using Eq. (7) to capture the overall trend of geographical clusters of ZIP codes based on their disaster impacts and recovery duration values.
\begin{equation}
\label{eq:7}
I\ =\ \frac{n}{\sum_{i}\sum_{j} w_{ij}}\frac{\sum_{i}\sum_{j}{w_{ij}z_iz_j}}{\sum_{i} z_i^2}      
\end{equation}
In Eq. (7), n represents the number of ZIP code; $w_{ij}$: the spatial weight of ZIP code j on ZIP code i; $z_i$/$z_j$: the deviations from the mean for ${DI}_i$/${DI}_j$ or ${RD}_i$/${RD}_j$.

\subsubsection{\textit{Local spatial autocorrelation}}
To determine local spatial autocorrelation, we used the local Moran’s I in the Local Indicators of Spatial Association (LISAs) to detect the clusters of ZIP codes with similar values of disaster impacts and recovery duration surrounded by high values (high-high) or low values (low-low), and also identify dissimilarities in clusters with opposite values (high-low and low-high). Using Eq. (8), we calculated the local Moran’s I for the two indices in each ZIP code and made the LISA maps for disaster impacts and recovery duration, accordingly.
\begin{equation}
\label{eq:8}
\begin{split}
I_i\ =\ \frac{z_i}{m_2}\sum_{j}{w_{ij}z_j} \\
where, m_2\ =\ \frac{\sum_{i} z_i^2}{n} 
\end{split}
\end{equation}
In Eq (8), $w_{ij}$, $z_i$/$z_j$ and n have the same meanings as those in Eq. (7).

\subsection{Spatial disparities of disaster impact and recovery duration}
By examining the spatial patterns of disaster impacts and recovery duration, we can observe disparities among various ZIP codes. In the next step, we examined the extent to which the disaster impacts and recovery duration varied across different ZIP codes with different extent of flood damage (measured based on the flood claims) and the sociodemographic characteristics of the residents (e.g., income and race). To do this, we performed the stepwise regression analysis with our calculated disaster impacts and recovery duration and their association with demographic and flood claim data across all ZIP codes in Harris County. Flood claim data refers to the insurance claim (number of claims) and reported loss (unit: US dollars) data, which comes from survey data from the Texas Governor’s Commission to Rebuild Texas (CRT) project. The demographics data for each ZIP code includes total population, gender (percentage of female population), ethnicity (percentage of white population), median of age, median of household income, and education (percentage of population with degree less than high school). 

To choose the best subset among all the predictors, a stepwise model selection was utilized. In this approach, the goal is to find the model which can properly fit the data and also control for the number of variables included in the model. Akaike information criterion (AIC) was used for the stepwise model selection. AIC criteria considers a penalty for including more variables in the model as shown in Eq. (9).
\begin{equation}
\label{eq:9}
AIC=-2\times\ln{(L)}+2\times(p)      
\end{equation}
Where, L refers to the likelihood of the fitted model, p is the number of parameters used in the model, and n is the sample size. After conducting the stepwise regression analysis, we used the variance inflation factor (VIF) to detect the multicollinearity exiting in the independent variables including demographic and claim data. VIF was calculated by Eq. (10).
\begin{equation}
\label{eq:10}
{VIF}_i\ =\ \frac{1}{1\ -\ R_i^2}      
\end{equation}
Where,  ${R_i}^2$ is the coefficient of determination in linear regression for variable i with all the other variables in the demographic and claim data, and its value ranges from 0 to 1. Then, we set the threshold of VIF as 5 to detect the multicollinearity issue among demographic and claim variables. 
%\bibliographystyle{unsrt}  

\section*{Acknowledgements}
The authors would like to acknowledge the funding support from the National Science Foundation CAREER Award under grant number 1846069. The authors would also like to acknowledge SafeGraph for providing the points-of-interest data. Any opinions, findings, conclusions, or recommendations expressed in this research are those of the authors and do not necessarily reflect the view of the funding agencies.

\section*{Author contributions}
F.Y., A.E., and A.M. conceived and designed the research. F.Y., A.E., and B.O. performed the numerical calculations. F.Y., A.E., and A.M. analyzed the results, and wrote and edited the manuscript.

\section*{References}
1.	Vugrin, E. D. et al. A Framework for Assessing the Resilience of Infrastructure and Economic Systems Acronyms, Abbreviations, and Initialisms. 2010 77–116

2.	Mostafavi, A., Abraham, D. M. \& Lee, J. System-of-systems approach for assessment of financial innovations in infrastructure. Built Environ. Proj. Asset Manag. 2, 250–265 (2012).

3.	Peacock, W. G., Van Zandt, S., Zhang, Y. \& Highfield, W. E. Inequities in long-term housing recovery after disasters. J. Am. Plan. Assoc. 80, 356–371 (2014).

4.	Liu, Y., Sui, Z., Kang, C. \& Gao, Y. Uncovering patterns of inter-urban trip and spatial interaction from social media check-in data. PLoS One 9, (2014).

5.	Markhvida, M., Walsh, B., Hallegatte, S. \& Baker, J. Quantification of disaster impacts through household well-being losses. Nat. Sustain. (2020).

6.	Dong, S., Esmalian, A., Farahmand, H. \& Mostafavi, A. An integrated physical-social analysis of disrupted access to critical facilities and community service-loss tolerance in urban flooding. Comput. Environ. Urban Syst. 80, (2020).

7.	Yuan, F., Li, M., \& Liu, R. Understanding the evolutions of public responses using social media: Hurricane Matthew case study. Int. J. Disaster Risk Reduct, 51, 101798 (2020).

8.	Zhai, W., Peng, Z. R. \& Yuan, F. Examine the effects of neighborhood equity on disaster situational awareness: Harness machine learning and geotagged Twitter data. Int. J. Disaster Risk Reduct. 48, 101611 (2020).

9.	Yuan, F., Li, M., Liu, R., Zhai, W., \& Qi, B. Social media for enhanced understanding of disaster resilience during Hurricane Florence. International Journal of Information Management, 57, 102289 (2021).

10. FEMA. National Mitigation Framework, Second Edition. (2016).

11.	Aghababaei, M., Koliou, M., Watson, M. \& Xiao, Y. Quantifying post-disaster business recovery through Bayesian methods. Struct. Infrastruct. Eng. 0, 1–19 (2020).

12.	Coleman, N., Esmalian, A. \& Mostafavi, A. Equitable Resilience in Infrastructure Systems: Empirical Assessment of Disparities in Hardship Experiences of Vulnerable Populations during Service Disruptions. Nat. Hazards Rev. (2019).

13.	Lazo, J. K., Bostrom, A., Morss, R. E., Demuth, J. L. \& Lazrus, H. Factors Affecting Hurricane Evacuation Intentions. Risk Anal. 35, 1837–1857 (2015).

14.	Martins, V. N., Nigg, J., Louis-Charles, H. M. \& Kendra, J. M. Household preparedness in an imminent disaster threat scenario: The case of superstorm sandy in New York City. Int. J. Disaster Risk Reduct. (2019). doi:10.1016/j.ijdrr.2018.11.003

15.	Rasoulkhani, K. \& Mostafavi, A. Resilience as an emergent property of human-infrastructure dynamics: A multi-agent simulation model for characterizing regime shifts and tipping point behaviors in infrastructure systems. PLoS One 13, e0207674 (2018).

16.	Fan, C., Jiang, Y. \& Mostafavi, A. Social Sensing in Disaster City Digital Twin: Integrated Textual–Visual–Geo Framework for Situational Awareness during Built Environment Disruptions. J. Manag. Eng. 36, 04020002 (2020).

17.	Chen, Z., Gong, Z., Yang, S., Ma, Q. \& Kan, C. Impact of extreme weather events on urban human flow: A perspective from location-based service data. Comput. Environ. Urban Syst. 83, 101520 (2020).

18.	Yuan, F., Liu, R., Mao, L. \& Li, M. Internet of people enabled framework for evaluating performance loss and resilience of urban critical infrastructures. Saf. Sci. 134, 105079 (2021).

19. Yuan, F., \& Liu, R. Mining social media data for rapid damage assessment during Hurricane Matthew: feasibility study. Journal of Computing in Civil Engineering, 34(3), 05020001, (2020).

20.Yuan, F., \& Liu, R. Feasibility study of using crowdsourcing to identify critical affected areas for rapid damage assessment: Hurricane Matthew case study. Int. J. Disaster Risk Reduct., 28, 758-767, (2018).

21.	Esmalian, A., Dong, S. \& Mostafavi, A. Susceptibility Curves for Humans: Empirical Survival Models for Determining Household-level Disturbances from Hazards-induced Infrastructure Service Disruptions. Sustain. Cities Soc. (2020).

22.	Esmalian, A., Dong, S., Coleman, N. \& Mostafavi, A. Determinants of risk disparity due to infrastructure service losses in disasters: a household service gap model. Risk Anal. (2019).

23.	Peacock, W. G. \& Ragsdale, A. K. Social systems, ecological networks and disasters: Toward a socio-political ecology of disasters. Hurric. Andrew Ethn. gender, Sociol. disasters 20–35 (1997).

24.	Knüsel, B. et al. Applying big data beyond small problems in climate research. Nat. Clim. Chang. 9, 196–202 (2019).

25.	Lenormand, M. et al. Influence of sociodemographics on human mobility. Sci. Rep. 5, 190–198 (2015).

26.	Hasan, S., Zhan, X. \& Ukkusuri, S. V. Understanding urban human activity and mobility patterns using large-scale location-based data from online social media. Proc. ACM SIGKDD Int. Conf. Knowl. Discov. Data Min. (2013). doi:10.1145/2505821.2505823

27.	Bagrow, J. P., Wang, D. \& Barabasi, A.-L. Collective Response of Human Populations to Large- Scale Emergencies. PLoS One 7, 1–10 (2012).

28.	Dong, X., Meyer, J., Shmueli, E., Bozkaya, B. \& Pentland, A. Methods for quantifying effects of social unrest using credit card transaction data. EPJ Data Sci. 7, 1–25 (2018).

29.	Chang, S. et al. Mobility network models of COVID-19 explain inequities and inform reopening. Nature (2020). doi:10.1038/s41586-020-2923-3

30.	Han, S. Y., Tsou, M., Knaap, E., Rey, S. \& Cao, G. How Do Cities Flow in an Emergency? Tracing Human Mobility Patterns during a Natural Disaster with Big Data and Geospatial Data Science. (2019).

31.	Juhász, L.\& Hochmair, H. Studying Spatial and Temporal Visitation Patterns of Points of Interest Using SafeGraph Data in Florida. GI Forum 1, 119–136 (2020).

32.	Alatrista-Salas, H., Gauthier, V., Nunez-del-Prado, M. \& Becker, M. Impact of natural disasters on consumer behavior: case of the 2017 El Nino phenomenon in Peru. 1–28 (2020).

33.	Martinez, E. A. et al. Measuring Economic Resilience to Natural Disasters with Big Economic Transaction Data. (2016).

34.	Yabe, T., Zhang, Y. \& Ukkusuri, S. Quantifying the Economic Impact of Extreme Shocks on Businesses using Human Mobility Data: a Bayesian Causal Inference Approach. arXiv:2004.11121 (2020).

35.	Statistical Altas. Overview of Harris County, Texas. (2020). 

36.	Visa. Visa Merchant Data Standards Manual. Journal of Innovative Image Processing 1, (2019).

37.	Nan, C. \& Sansavini, G. A quantitative method for assessing resilience of interdependent infrastructures. Reliab. Eng. Syst. Saf. 157, 35–53 (2017).

38.	Anselin, L. \& Smirnov, O. Efficient algorithms for constructing proper higher order spatial LAG operators. J. Reg. Sci. 36, 67–89 (1996).

39.	Ouyang, M. \& Fang, Y. A Mathematical Framework to Optimize Critical Infrastructure Resilience against Intentional Attacks. Comput. Civ. Infrastruct. Eng. 32, 909–929 (2017).

40.	Guidotti, R. et al. Modeling the resilience of critical infrastructure: the role of network dependencies. Sustain. Resilient Infrastruct. 1, 153–168 (2016).

41.	Claeys, A.-S. \& Cauberghe, V. What makes crisis response strategies work? The impact of crisis involvement and message framing. J. Bus. Res. 67, 182–189 (2014).

42.	Podesta, C., Coleman, N., Esmalian, A., Yuan, F. \& Mostafavi, A. Quantifying community resilience based on fluctuations in visits to point-of-interest from digital trace data. arXiv 1–23 (2020).

\section*{Supplemental Information}
\appendix
\counterwithin{figure}{section}

% \usepackage{multirow}
\begin{table}[h]
\renewcommand\thetable{S-1} 
   \centering
   \caption{MCCs of business sectors}
\begin{tabular}{ll}
\hline
Business sector                                                                             & MCCs (merchant category codes)                                                                                                                                                                                                                                                                                                                                                                                                                                                                                                                                                     \\ \hline
Auto                                                                                        & 5511, 5521,   5532, 5533, 5541, 5542, 5551, 5561, 5571, 5592, 5599.                                                                                                                                                                                                                                                                                                                                                                                                                                                                                       \\
Cloth                                                                                       & 5611, 5621, 5631,   5641, 5651, 5655, 5661, 5681, 5691, 5697, 5698, 5699.                                                                                                                                                                                                                                                                                                                                                                                                                                                                                 \\
Drug store                                                                                  & 5122, 5912.                                                                                                                                                                                                                                                                                                                                                                                                                                                                                                                                               \\
Financial sector                                                                            & 6010, 6011, 6012,   6051, 6211, 6513, 6540, 9222, 9223, 9311.                                                                                                                                                                                                                                                                                                                                                                                                                                                                                             \\
\begin{tabular}[c]{@{}l@{}}Governmental   \\ sector\end{tabular}                            & 9399, 9402.                                                                                                                                                                                                                                                                                                                                                                                                                                                                                                                                               \\
Grocery                                                                                     & 5199, 5300, 5309,   5310, 5311, 5331, 5399, 5411, 5422, 5441, 5451, 5462, 5499.                                                                                                                                                                                                                                                                                                                                                                                                                                                                           \\
Health care                                                                                 & 8011, 8062, 8099.                                                                                                                                                                                                                                                                                                                                                                                                                                                                                                                                         \\
Home supply                                                                                 & 5712, 5713, 5714,   5718, 5719, 5722, 5732.                                                                                                                                                                                                                                                                                                                                                                                                                                                                                                               \\
\begin{tabular}[c]{@{}l@{}}Hotel, inns and   \\ lodging\end{tabular}                        & \begin{tabular}[c]{@{}l@{}}3501, 3502, 3503,   3504, 3508, 3509, 3510, 3512, 3513, 3515, 3516, 3528, 3530, 3535, \\ 3548, 3549,   3559, 3562, 3581, 3589, 3590, 3604, 3607, 3615, 3617, 3619, 3621, 3627, \\ 3629,   3637, 3638, 3640, 3641, 3644, 3649, 3650, 3652, 3655, 3663, 3665, 3687, 3690, \\ 3692, 3695, 3697, 3700, 3703, 3707, 3709, 3715, 3719, 3722, 3726, 3740, 3743, 3750, \\ 3751, 3778, 3780, 3791, 3812, 3814, 3826, 3829.\end{tabular}                                                                                                 \\
Insurance                                                                                   & 5960, 6300.                                                                                                                                                                                                                                                                                                                                                                                                                                                                                                                                               \\
\begin{tabular}[c]{@{}l@{}}Internet \&\\ Telecommunication\end{tabular}                     & 4812, 4813, 4814,   4816, 4821, 4829, 4899.                                                                                                                                                                                                                                                                                                                                                                                                                                                                                                               \\
Marketing                                                                                   & 5964, 5965, 5966,   5967, 5968, 5969.                                                                                                                                                                                                                                                                                                                                                                                                                                                                                                                     \\
Organization                                                                                & 8398, 8641, 8651,   8661, 8699.                                                                                                                                                                                                                                                                                                                                                                                                                                                                                                                           \\
Recreation                                                                                  & \begin{tabular}[c]{@{}l@{}}5815, 5816, 5817,   5818, 7012, 7032, 7033, 7800, 7801, 7829, 7832, 7841, 7911, 7922, \\ 7929, 7933, 7941, 7991, 7992, 7993, 7994, 7995, 7996, 7997, 7998, 7999, 8675.\end{tabular}                                                                                                                                                                                                                                                                                                                                            \\
Restaurant                                                                                  & 5811, 5812, 5813,   5814.                                                                                                                                                                                                                                                                                                                                                                                                                                                                                                                                 \\
Retail                                                                                      & \begin{tabular}[c]{@{}l@{}}5013, 5021, 5045,   5046, 5047, 5051, 5085, 5094, 5099, 5111, 5131, 5137, 5139, 5169, \\ 5172, 5192, 5193, 5198, 5200, 5231, 5251, 5261, 5271, 5921, 5931, 5932, 5933, 5935, \\ 5937, 5940, 5941, 5942, 5943, 5944, 5945, 5946, 5947, 5948, 5949, 5950, 5970, 5971,   \\ 5972, 5973, 5976, 5992, 5993, 5995, 5996, 5997, 5999.\end{tabular}                                                                                                                                                                                    \\
Schools                                                                                     & 8211, 8220, 8221,   8244, 8249, 8299.                                                                                                                                                                                                                                                                                                                                                                                                                                                                                                                     \\
Self-care                                                                                   & 8021, 8041, 8042,   8043, 8050, 8071.                                                                                                                                                                                                                                                                                                                                                                                                                                                                                                                     \\
Service                                                                                     & \begin{tabular}[c]{@{}l@{}}742, 763, 780, 1520,1711, 1731, 1740, 1750, 1761, 1771, 1799, 2741, 2791, 2842, \\ 4225, 4722, 5039, 5044, 5065, 5072, 5074, 5211, 5733, 5734, 5735, 5963, 5977, \\ 5994, 7011, 7210, 7211, 7216, 7217, 7221, 7230, 7251, 7273, 7276, 7277, 7278, \\ 7296, 7297, 7298, 7299, 7311, 7321, 7333, 7338, 7339, 7342, 7349, 7361, 7372, \\ 7375, 7379, 7392,   7393, 7394, 7395, 7399, 7531, 7534, 7535, 7538, 7542, 7549, \\ 7622, 7623, 7629,   7631, 7641, 7692, 7699, 8111, 8351, 8734, 8911, 8931, 8999, \\ 9211.\end{tabular} \\
Transportation                                                                              & \begin{tabular}[c]{@{}l@{}}3000, 3001, 3031,   3032, 3058, 3059, 3066, 3132, 3174, 3196, 3211, 3246, 3256, 3260, \\ 3351, 3357,   3359, 3366, 3368, 3389, 3390, 3395, 3405, 4111, 4112, 4119, 4121, 4131, \\ 4214, 4215, 4411, 4457, 4468, 4511, 4784, 4789, 7512, 7513, 7519, 7523.\end{tabular}                                                                                                                                                                                                                                                         \\
\begin{tabular}[c]{@{}l@{}}Utilities –   Electric, Gas, \\ Water, and Sanitary\end{tabular} & 4900, 5983.                                                                                                                                                                                                                                                                                                                                                                                                                                                                                                                                               \\ \hline
\end{tabular}
\end{table}

\begin{table}[]
\renewcommand\thetable{S-2} 
   \centering
   \caption{Gaussian regression analysis for disaster impact of auto}
\begin{tabular}{lllll}
\hline
Variables        & Coefficient & Std. Error & P-values        & VIF  \\ \hline
Intercept        & -4.20 e-02  & 5.18 e-02  & 0.420           & -    \\
Number of claims & 2.40        & 0.96       & 0.013           & 1.07 \\
Total population & -9.73 e-07  & -6.47 e-07 & .1353           & 1.03 \\
White percentage & 0.18        & 6.96 e-02  & 0.011           & 1.14 \\
Income           & 3.46 e-06   & 6.45 e-07  & \textless{}.001 & 1.20 \\ \hline
\end{tabular}
\end{table}

\begin{table}[]
\renewcommand\thetable{S-3} 
   \centering
   \caption{Gaussian regression analysis for disaster impact of cloth}
\begin{tabular}{lllll}
\hline
Variables  & Coefficient & Std. Error & P-values        & VIF  \\ \hline
Intercept  & 0.70        & 7.67 e-02  & \textless{}.001 & -    \\
White (\%) & 0.32        & 0.12       & 0.008           & 1.13 \\
Income     & -3.63 e-06  & 1.10 e-06  & 0.001           & 1.13 \\ \hline
\end{tabular}
\end{table}

\begin{table}[]
\renewcommand\thetable{S-4} 
   \centering
   \caption{Gaussian regression analysis for disaster impact of grocery}
\begin{tabular}{lllll}
\hline
Variables        & Coefficient & Std. Error & P-values         & VIF  \\ \hline
Intercept        & -0.18       & 0.18       & 0.330            & -    \\
Total population & -1.55 e-06  & 8.18 e-07  & 0.060            & 1.17 \\
Age              & 9.34 e-03   & 5.02 e-03  & 0.066            & 1.76 \\
Education (\%)   & 0.80        & 0.27       & 0.004            & 2.50 \\
Income           & 5.62 e-06   & 1.15 e-06  & \textless{}0.001 & 2.76 \\ \hline
\end{tabular}
\end{table}

\begin{table}[]
\renewcommand\thetable{S-5} 
   \centering
   \caption{Gaussian regression analysis for disaster impact of home supply}
\begin{tabular}{lllll}
\hline
Variables      & Coefficient & Std. Error & P-values         & VIF  \\ \hline
Intercept      & 1.57 e-02   & 6.44 e-02  & 0.808            & -    \\
Education (\%) & 0.90        & 0.26       & \textless{}0.001 & 2.15 \\
Income         & 7.18 e-06   & 1.04 e-06  & \textless{}0.001 & 2.15 \\ \hline
\end{tabular}
\end{table}

\begin{table}[]
\renewcommand\thetable{S-6} 
   \centering
   \caption{Gaussian regression analysis for disaster impact of IT}
\begin{tabular}{lllll}
\hline
Variables        & Coefficient & Std. Error & P-values         & VIF  \\ \hline
Intercept        & 0.30        & 7.06 e-02  & \textless{}0.001 & -    \\
Total population & -2.90 e-06  & 8.86 e-07  & 0.001            & 1.02 \\
White (\%)       & 0.16        & 9.55 e-02  & 0.090            & 1.13 \\
Income           & 4.62 e-06   & 8.70 e-07  & \textless{}0.001 & 1.15 \\ \hline
\end{tabular}
\end{table}

\begin{table}[]
\renewcommand\thetable{S-7} 
   \centering
   \caption{Gaussian regression analysis for disaster impact of marketing}
\begin{tabular}{lllll}
\hline
Variables     & Coefficient & Std. Error & P-values         & VIF  \\ \hline
Intercept     & 0.22        & 0.10       & 0.033            & -    \\
Reported loss & 3.38 e-05   & 2.31 e-05  & 0.100            & 1.01 \\
White (\%)    & 0.67        & 0.15       & \textless{}0.001 & 1.01 \\ \hline
\end{tabular}
\end{table}

\begin{table}[]
\renewcommand\thetable{S-8} 
   \centering
   \caption{Gaussian regression analysis for disaster impact of recreation}
\begin{tabular}{lllll}
\hline
Variables        & Coefficient & Std. Error & P-values         & VIF  \\ \hline
Intercept        & 0.42        & 0.11       & \textless{}.001  & -    \\
Number of claims & 4.14        & 2.07       & 0.047            & 1.03 \\
Total population & -7.14 e-06  & 1.42 e-06  & \textless{}0.001 & 1.01 \\
White (\%)       & 0.55        & 0.15       & \textless{}0.001 & 1.02 \\ \hline
\end{tabular}
\end{table}

\begin{table}[]
\renewcommand\thetable{S-9} 
   \centering
   \caption{Gaussian regression analysis for disaster impact of restaurant}
\begin{tabular}{lllll}
\hline
Variables        & Coefficient & Std. Error & P-values & VIF  \\ \hline
Intercept        & 0.15        & 0.11       & 0.163    & -    \\
Number of claims & 2.28        & 0.74       & 0.003    & 1.09 \\
Age              & 4.71 e-03   & 3.30 e-03  & 0.157    & 1.77 \\
Education (\%)   & 0.29        & 0.17       & 0.088    & 2.21 \\
Income           & 2.38 e-06   & 7.36 e-07  & 0.002    & 2.62 \\ \hline
\end{tabular}
\end{table}

\begin{table}[]
\renewcommand\thetable{S-10} 
   \centering
   \caption{Gaussian regression analysis for disaster impact of retail}
\begin{tabular}{lllll}
\hline
Variables        & Coefficient & Std. Error & P-values         & VIF  \\ \hline
Intercept        & 0.32        & 0.13       & 0.018            & -    \\
Reported loss    & -2.81 e-05  & 1.96 e-05  & 0.155            & 1.05 \\
Total population & -3.39 e-06  & 1.30 e-06  & 0.010            & 1.13 \\
Education (\%)   & 0.97        & 0.44       & 0.028            & 2.42 \\
Income           & 7.52 e-06   & 1.76 e-06  & \textless{}0.001 & 2.42 \\ \hline
\end{tabular}
\end{table}

\begin{table}[]
\renewcommand\thetable{S-11} 
   \centering
   \caption{Gaussian regression analysis for disaster impact of service}
\begin{tabular}{lllll}
\hline
Variables      & Coefficient & Std. Error & P-values         & VIF  \\ \hline
Intercept      & -.19        & 0.11       & 0.080            & -    \\
White (\%)     & 0.27        & 0.13       & 0.038            & 1.20 \\
Education (\%) & 1.46        & 0.41       & \textless{}0.001 & 2.28 \\
Income         & 9.10 e-06   & 1.70 e-06  & \textless{}0.001 & 2.55 \\ \hline
\end{tabular}
\end{table}

\begin{table}[]
\renewcommand\thetable{S-12} 
   \centering
   \caption{Gaussian regression analysis for disaster impact of transportation}
\begin{tabular}{lllll}
\hline
Variables        & Coefficient & Std. Error & P-values         & VIF  \\ \hline
Intercept        & 0.33        & 0.12       & 0.007            & -    \\
Total population & -4.44 e-06  & 1.55 e-06  & 0.004            & 1.02 \\
White (\%)       & 0.68        & 0.17       & \textless{}0.001 & 1.13 \\
Income           & 5.31 e-06   & 1.52 e-06  & \textless{}0.001 & 1.15 \\ \hline
\end{tabular}
\end{table}

\begin{table}[]
\renewcommand\thetable{S-13} 
   \centering
   \caption{Poisson regression analysis for recovery duration of auto}
\begin{tabular}{lllll}
\hline
Variables      & Coefficient & Std. Error & P-values         & VIF  \\ \hline
Intercept      & 0.76        & 0.45       & 0.091            & -    \\
Reported loss  & -9.75 e-05  & 3.51 e-05  & 0.005            & 1.12 \\
White (\%)     & 1.51        & 0.22       & \textless{}0.001 & 1.13 \\
Age            & 2.57 e-02   & 1.30 e-02  & 0.047            & 1.88 \\
Education (\%) & -2.98       & 0.67       & \textless{}0.001 & 2.31 \\
Income         & -1.24 e-05  & 3.00 e-06  & \textless{}0.001 & 2.67 \\ \hline
\end{tabular}
\end{table}

\begin{table}[]
\renewcommand\thetable{S-14} 
   \centering
   \caption{Poisson regression analysis for recovery duration of cloth}
\begin{tabular}{lllll}
\hline
Variables        & Coefficient & Std. Error & P-values         & VIF  \\ \hline
Intercept        & 3.54        & 0.29       & \textless{}0.001 & -    \\
Number of claims & 4.56        & 2.13       & \textless{}0.032 & 1.10 \\
Total population & 6.03 e-06   & 1.27 e-06  & \textless{}0.001 & 1.03 \\
Age              & -3.77 e-02  & 9.50 e-03  & \textless{}0.001 & 1.68 \\
Income           & -3.63 e-06  & 1.86 e-06  & 0.0513           & 1.66 \\ \hline
\end{tabular}
\end{table}

\begin{table}[]
\renewcommand\thetable{S-15} 
   \centering
   \caption{Poisson regression analysis for recovery duration of grocery}
\begin{tabular}{lllll}
\hline
Variables      & Coefficient & Std. Error & P-values & VIF  \\ \hline
Intercept      & 1.23        & 0.40       & 0.002    & -    \\
Age            & 1.55 e-02   & 1.07 e-02  & 0.148    & 1.42 \\
Education (\%) & 1.00        & 0.48       & 0.037    & 1.42 \\ \hline
\end{tabular}
\end{table}

\begin{table}[]
\renewcommand\thetable{S-16} 
   \centering
   \caption{Poisson regression analysis for recovery duration of home supply}
\begin{tabular}{lllll}
\hline
Variables        & Coefficient & Std. Error & P-values         & VIF  \\ \hline
Intercept        & 2.13        & 0.15       & \textless{}0.001 & -    \\
Number of claims & -7.07       & 3.16       & 0.026            & 1.06 \\
Total population & 5.70 e-06   & 1.62 e-06  & \textless{}0.001 & 1.05 \\
White (\%)       & -0.64       & 0.17       & \textless{}0.001 & 1.03 \\
Education (\%)   & 0.81        & 0.42       & 0.053            & 1.05 \\ \hline
\end{tabular}
\end{table}

\begin{table}[]
\renewcommand\thetable{S-17} 
   \centering
   \caption{Poisson regression analysis for recovery duration of IT}
\begin{tabular}{lllll}
\hline
Variables        & Coefficient & Std. Error & P-values         & VIF  \\ \hline
Intercept        & 2.15        & 0.29       & \textless{}0.001 & -    \\
Total population & 4.70 e-06   & 1.25 e-06  & \textless{}0.001 & 1.03 \\
White (\%)       & -0.33       & 0.14       & 0.022            & 1.14 \\
Age              & 1.43 e-02   & 8.84 e-03  & 0.106            & 1.67 \\
Income           & -3.19 e-06  & 1.77 e-06  & 0.072            & 1.82 \\ \hline
\end{tabular}
\end{table}

\begin{table}[]
\renewcommand\thetable{S-18} 
   \centering
   \caption{Poisson regression analysis for recovery duration of marketing}
\begin{tabular}{lllll}
\hline
Variables        & Coefficient & Std. Error & P-values         & VIF   \\ \hline
Intercept        & 1.31        & 0.47       & 0.005            & -     \\
Number of claims & -5.60       & 3.22       & 0.082            & 1.09  \\
Total population & 2.86 e-06   & 1.83 e-06  & 0.119            & 1.191 \\
White (\%)       & 0.81        & 0.21       & \textless{}0.001 & 1.17  \\
Age              & 1.83 e-02   & 1.24 e-02  & 0.141            & 1.85  \\
Education (\%)   & -2.43       & 0.69       & \textless{}0.001 & 2.60  \\
Income           & -9.94 e-06  & -3.06 e-06 & 0.001            & 2.90  \\ \hline
\end{tabular}
\end{table}

\begin{table}[]
\renewcommand\thetable{S-19} 
   \centering
   \caption{Poisson regression analysis for recovery duration of recreation}
\begin{tabular}{lllll}
\hline
Variables        & Coefficient & Std. Error & P-values         & VIF  \\ \hline
Intercept        & 1.73        & 0.15       & \textless{}0.001 & -    \\
Number of claims & 3.57        & 2.31       & 0.123            & 1.02 \\
Total population & -8.17 e-06  & -2.04 e-06 & \textless{}0.001 & 1.02 \\
White (\%)       & 0.56        & 0.19       & 0.004            & 1.02 \\ \hline
\end{tabular}
\end{table}

\begin{table}[]
\renewcommand\thetable{S-20} 
   \centering
   \caption{Poisson regression analysis for recovery duration of restaurant}
\begin{tabular}{lllll}
\hline
Variables     & Coefficient & Std. Error & P-values         & VIF  \\ \hline
Intercept     & 1.88        & 0.11       & \textless{}0.001 & -    \\
Reported loss & -7.81 e-05  & 2.62 e-05  & 0.003            & 1.02 \\
White (\%)    & 0.26        & 0.17       & 0.129            & 1.15 \\
Income        & 6.10 e-06   & 1.36 e-06  & \textless{}0.001 & 1.16 \\ \hline
\end{tabular}
\end{table}

\begin{table}[]
\renewcommand\thetable{S-21} 
   \centering
   \caption{Poisson regression analysis for recovery duration of retail}
\begin{tabular}{lllll}
\hline
Variables        & Coefficient & Std. Error & P-values         & VIF  \\ \hline
Intercept        & 1.17        & 0.14       & \textless{}0.001 & -    \\
Total population & 5.19 e-06   & 1.54 e-06  & 0.001            & 1.02 \\
White (\%)       & 0.51        & 0.18       & 0.004            & 1.10 \\
Income           & -5.52 e-06  & 1.76 e-06  & 0.002            & 1.10 \\ \hline
\end{tabular}
\end{table}

\begin{table}[]
\renewcommand\thetable{S-22} 
   \centering
   \caption{Poisson regression analysis for recovery duration of service}
\begin{tabular}{lllll}
\hline
Variables        & Coefficient & Std. Error & P-values         & VIF  \\ \hline
Intercept        & 1.20        & 0.15       & \textless{}0.001 & -    \\
Total population & 8.45 e-06   & 1.55 e-06  & \textless{}0.001 & 1.02 \\
White (\%)       & 0.66        & 0.19       & \textless{}0.001 & 1.02 \\ \hline
\end{tabular}
\end{table}

\begin{table}[]
\renewcommand\thetable{S-23} 
   \centering
   \caption{Poisson regression analysis for recovery duration of transportation}
\begin{tabular}{lllll}
\hline
Variables        & Coefficient & Std. Error & P-values         & VIF  \\ \hline
Intercept        & 2.58        & 0.39       & \textless{}0.001 & -    \\
Total population & -3.34 e-06  & 1.73 e-06  & 0.054            & 1.16 \\
White (\%)       & 0.92        & 0.18       & \textless{}0.001 & 1.15 \\
Age              & -1.91 e-02  & 1.04 e-02  & 0.066            & 1.79 \\
Education (\%)   & -0.83       & 0.57       & 0.145            & 2.68 \\
Income           & -3.99 e-06  & -2.53 e-06 & 0.115            & 3.03 \\ \hline
\end{tabular}
\end{table}

\begin{figure}[ht]
\renewcommand\thefigure{S-1-9} 
\centering
\includegraphics[width=\linewidth]{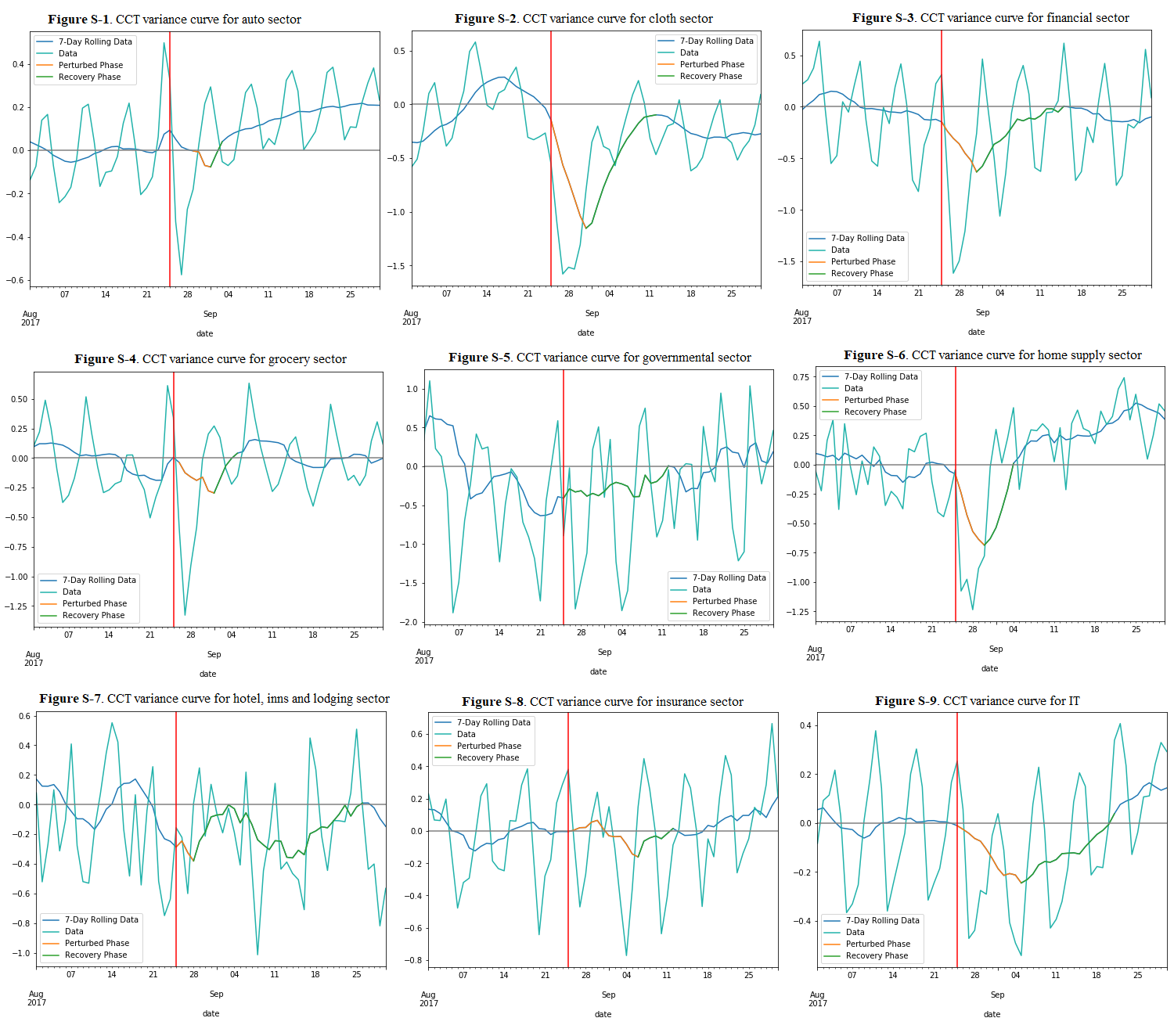}
\caption{CCT variance curves for auto, cloth, financial, grocery, governmental, home supply, hotel, inns and lodging, insurance and IT sectors}
\end{figure}

\begin{figure}[ht]
\renewcommand\thefigure{S-10-18} 
\centering
\includegraphics[width=\linewidth]{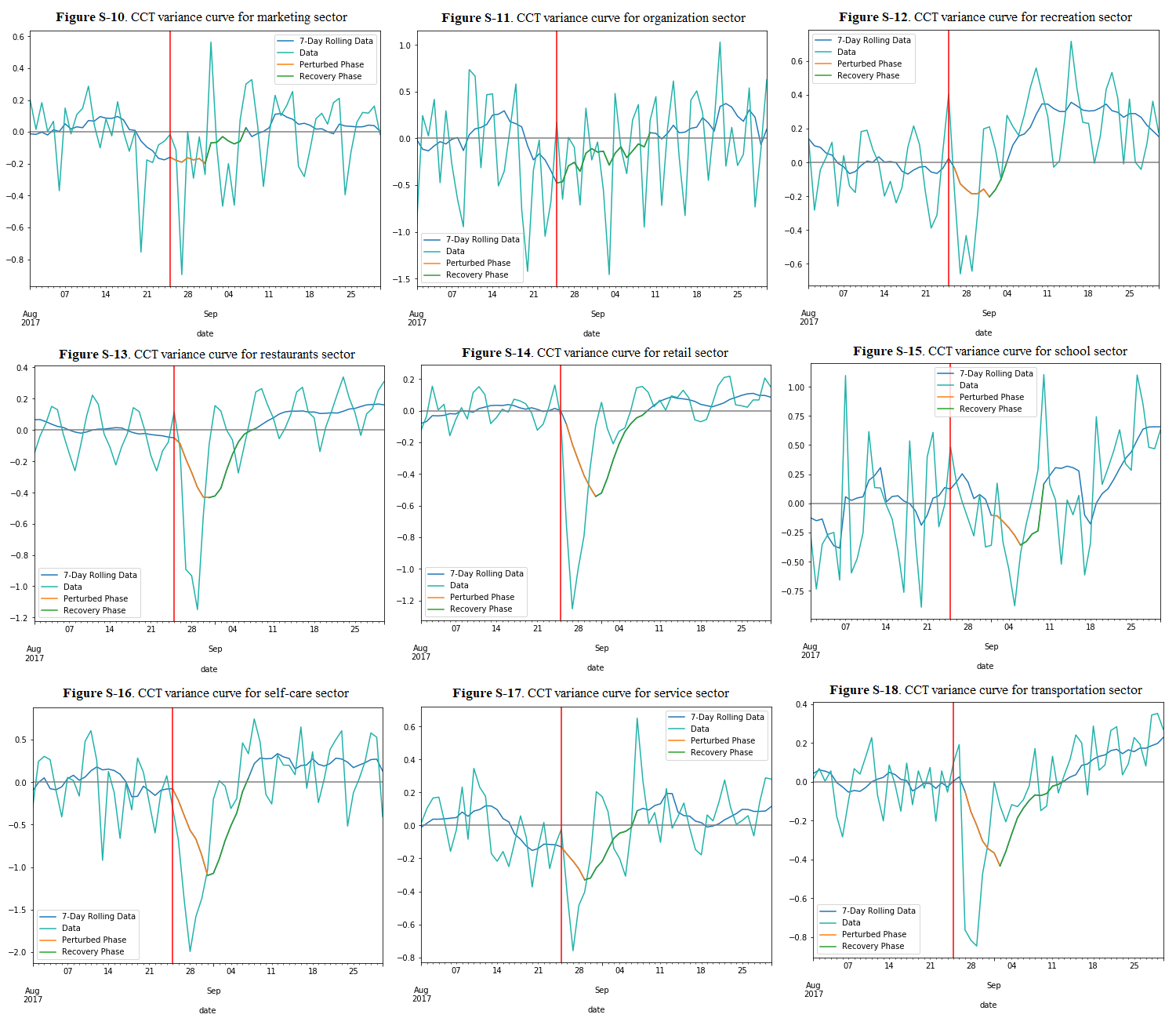}
\caption{CCT variance curves for marketing, organization, recreation, restaurant, retail, school, self-care, service and transportation sectors}
\end{figure}

\begin{figure}[ht]
\renewcommand\thefigure{S-19-22} 
\centering
\includegraphics[width=\linewidth]{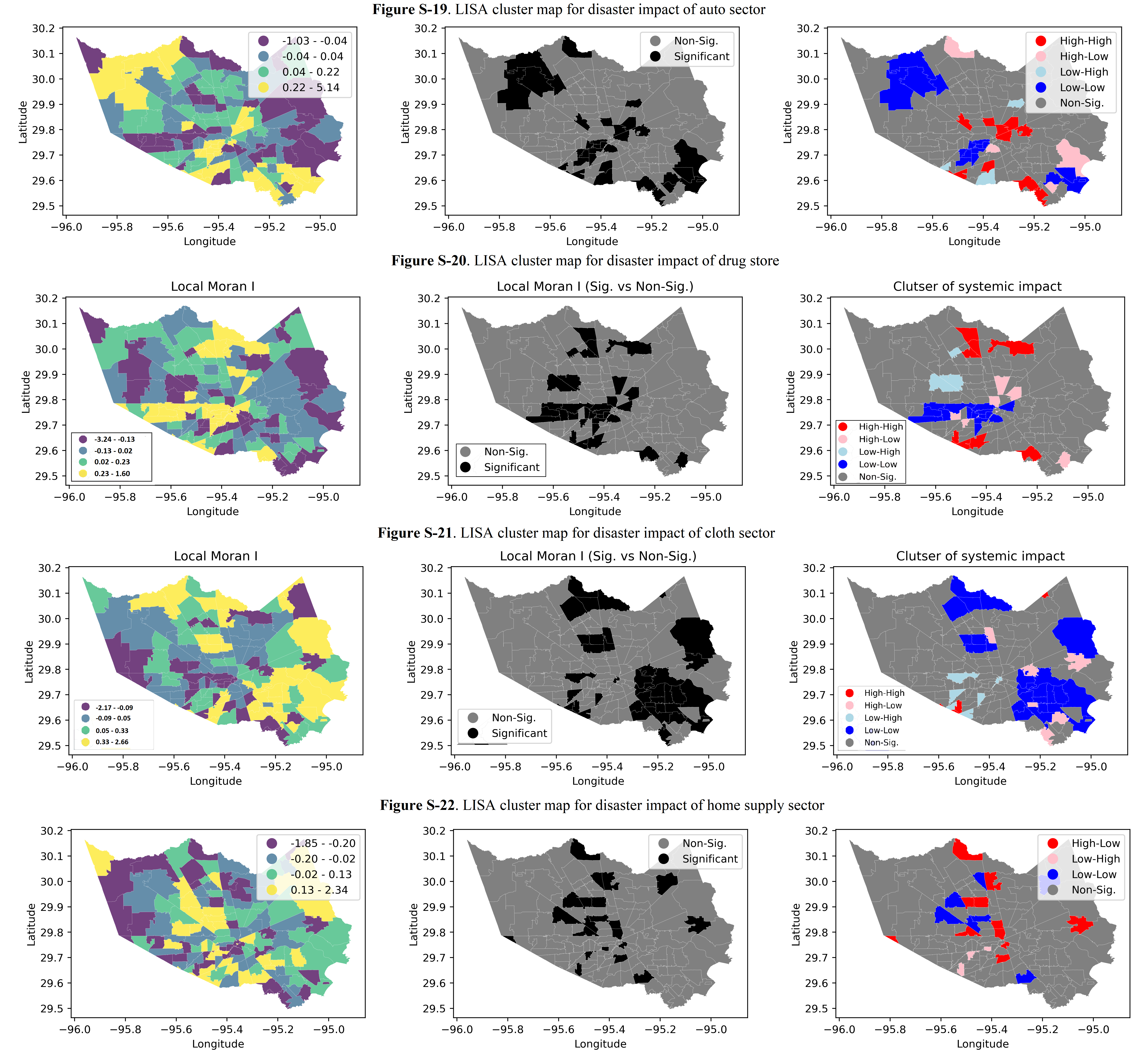}
\caption{LISA cluster maps for disaster impact of auto, drugstore, cloth and home supply sectors}
\end{figure}

\begin{figure}[ht]
\renewcommand\thefigure{S-23-26} 
\centering
\includegraphics[width=\linewidth]{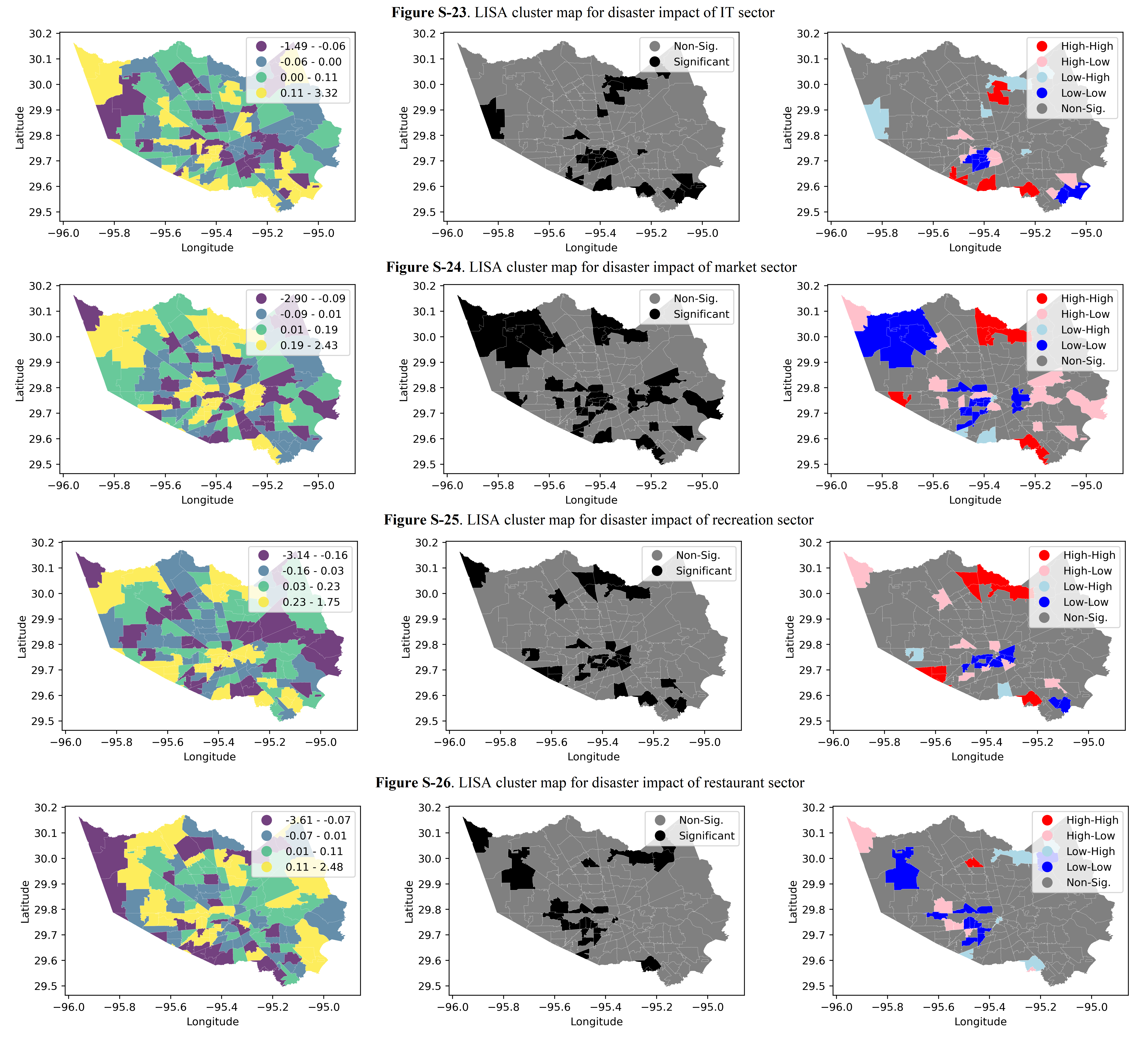}   
\caption{LISA cluster maps for disaster impact of IT, market, recreation and restaurant sectors}
\end{figure}

\begin{figure}[ht]
\renewcommand\thefigure{S-27-29} 
\centering
\includegraphics[width=\linewidth]{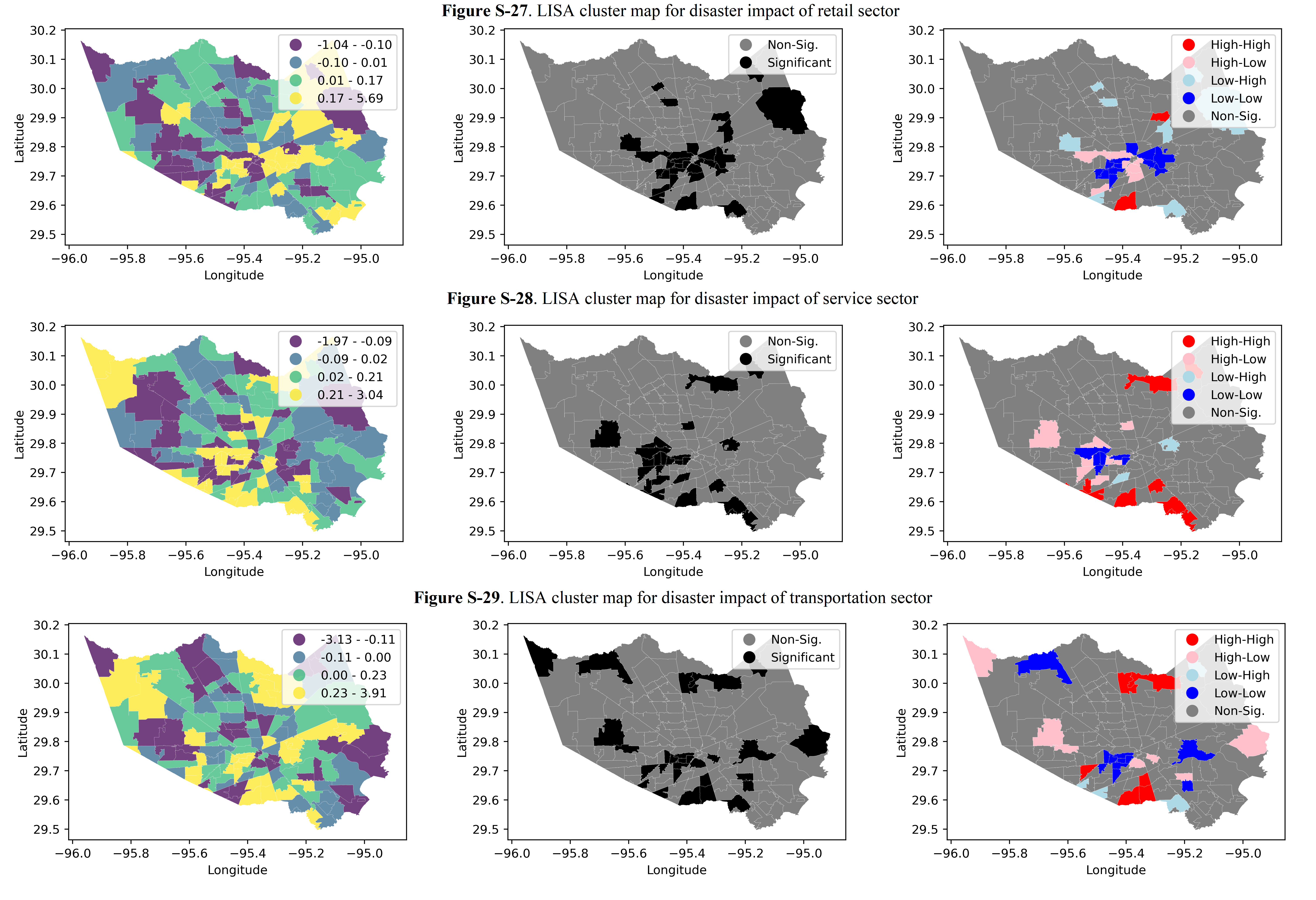}
\caption{LISA cluster maps for disaster impact of retail, service and transportation sectors}
\end{figure}

\begin{figure}[ht]
\renewcommand\thefigure{S-30-33} 
\centering
\includegraphics[width=\linewidth]{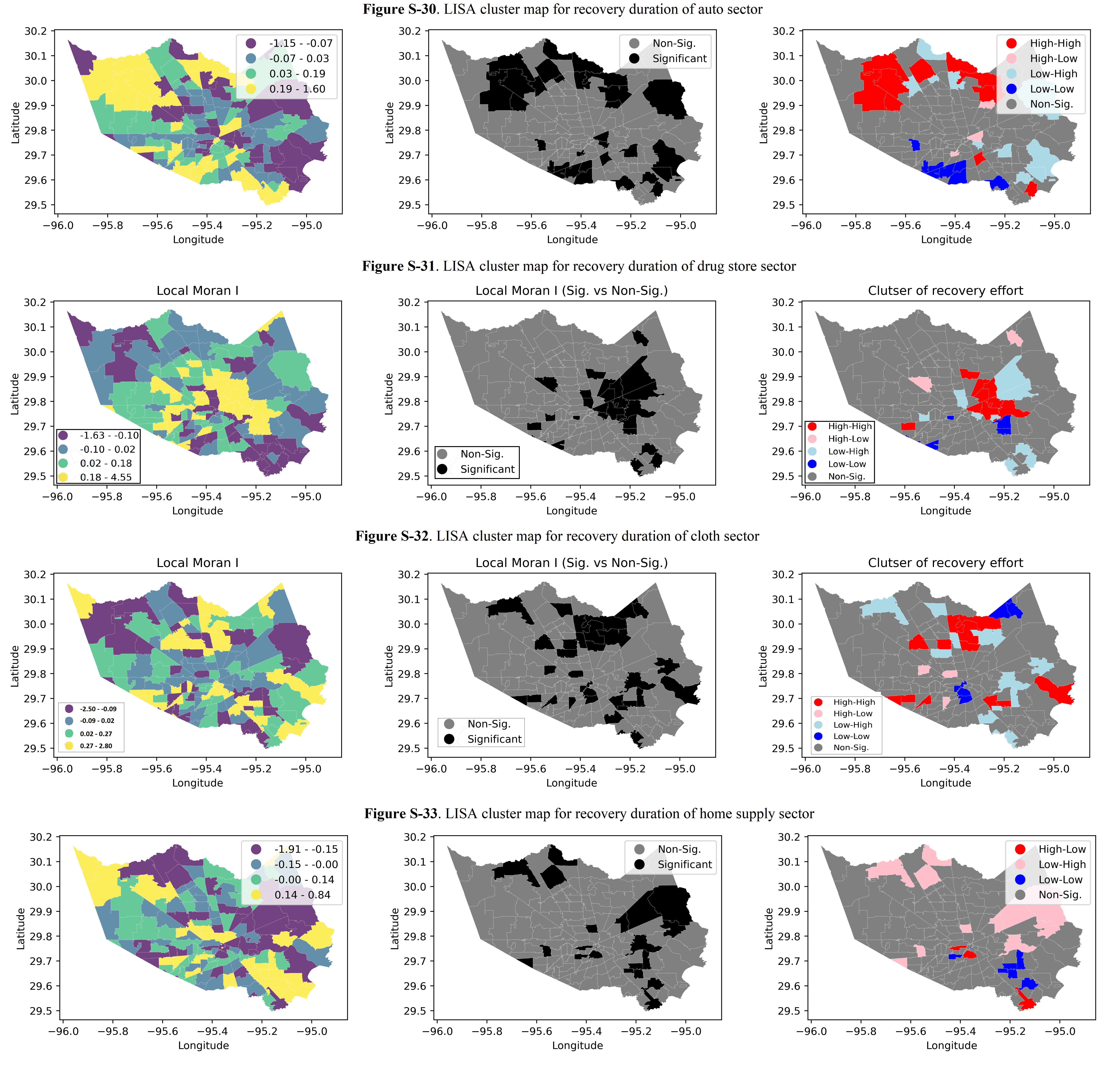}
\caption{LISA cluster maps for recovery duration of auto, drugstore, cloth and home suppl sectors}
\end{figure}

\begin{figure}[ht]
\renewcommand\thefigure{S-34-37} 
\centering
\includegraphics[width=\linewidth]{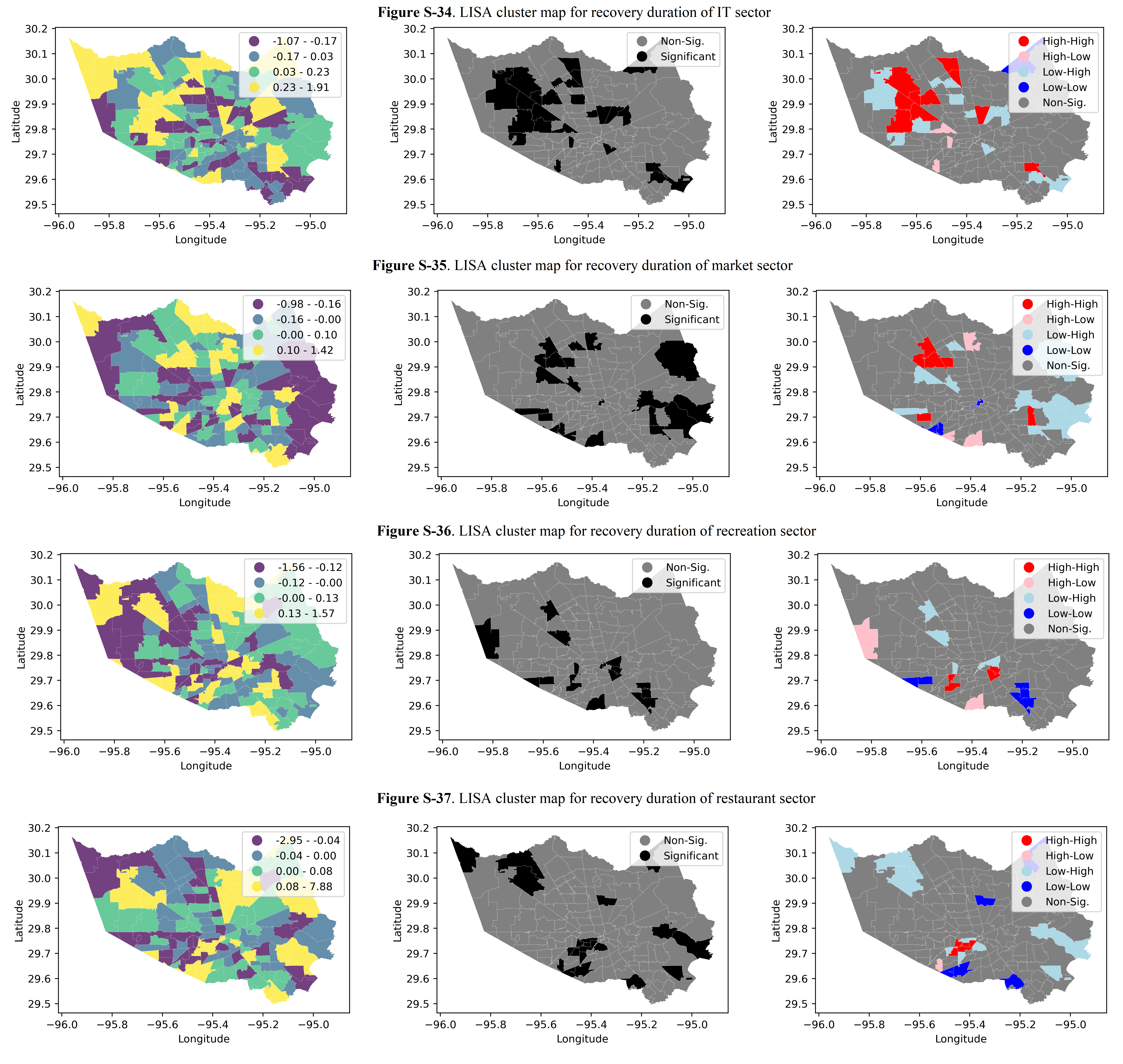}   
\caption{LISA cluster maps for recovery duration of IT, market, recreation and restaurant sectors}
\end{figure}

\begin{figure}[ht]
\renewcommand\thefigure{S-38-40} 
\centering
\includegraphics[width=\linewidth]{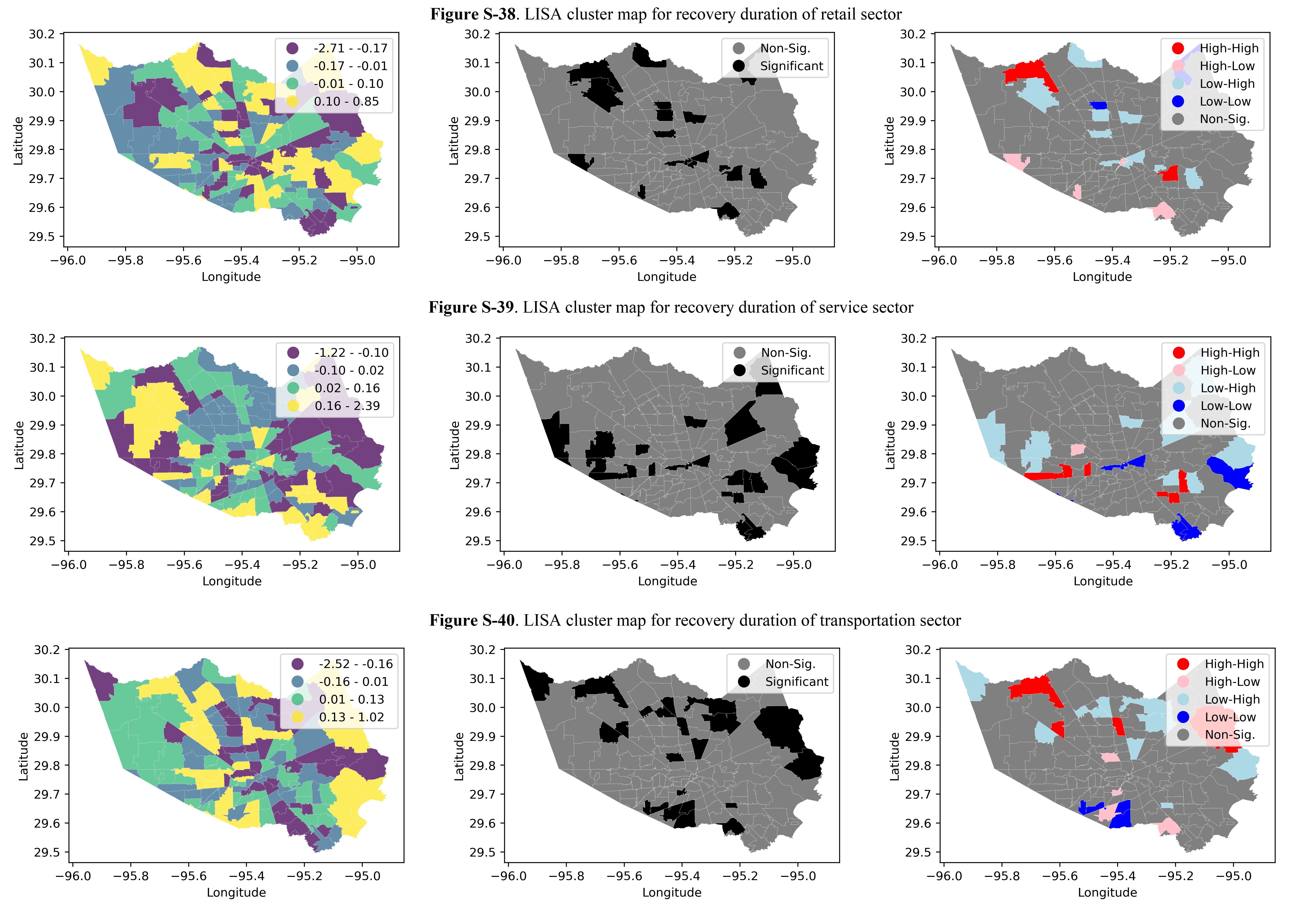}
\caption{LISA cluster maps for recovery duration of retail, service and transportation sectors}
\end{figure}

\begin{figure}[ht]
\renewcommand\thefigure{S-41} 
\centering
\includegraphics[width=\linewidth]{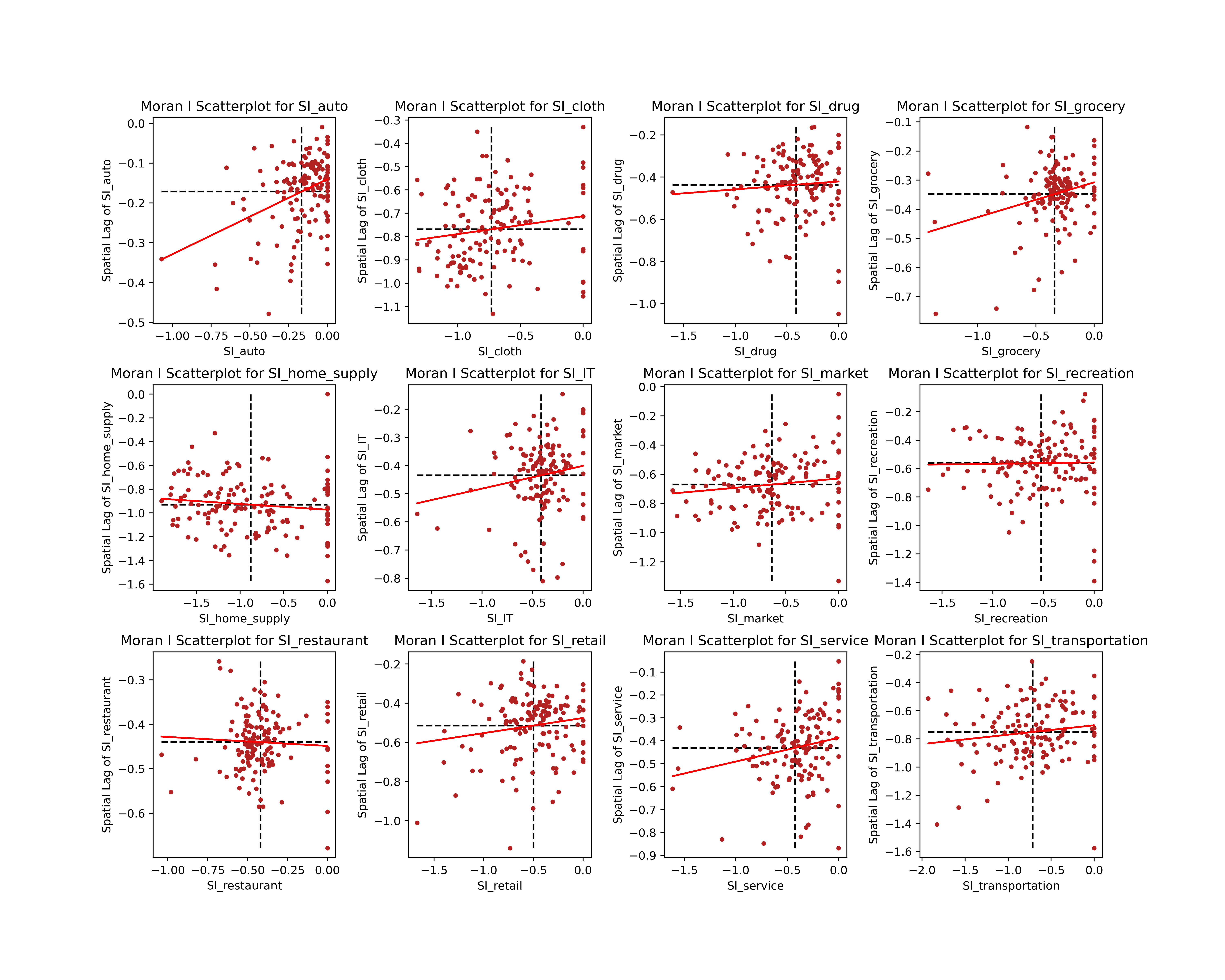}
\caption{Global Moran’s I scatterplot for disaster impact (DI: disaster impact). In Figures S-41 and S-42, the horizontal axis represents the values of disaster impact and recovery duration, while the vertical axis is for their spatial lag values. The horizontal dash line remarks the location of mean of the spatial lag value, while the vertical dash line marks the mean of disaster impact and recovery duration. The red line is the best linear fit for the scatterplot and its slope is the global Moran’s I.}
\end{figure}

\begin{figure}[ht]
\renewcommand\thefigure{S-42} 
\centering
\includegraphics[width=\linewidth]{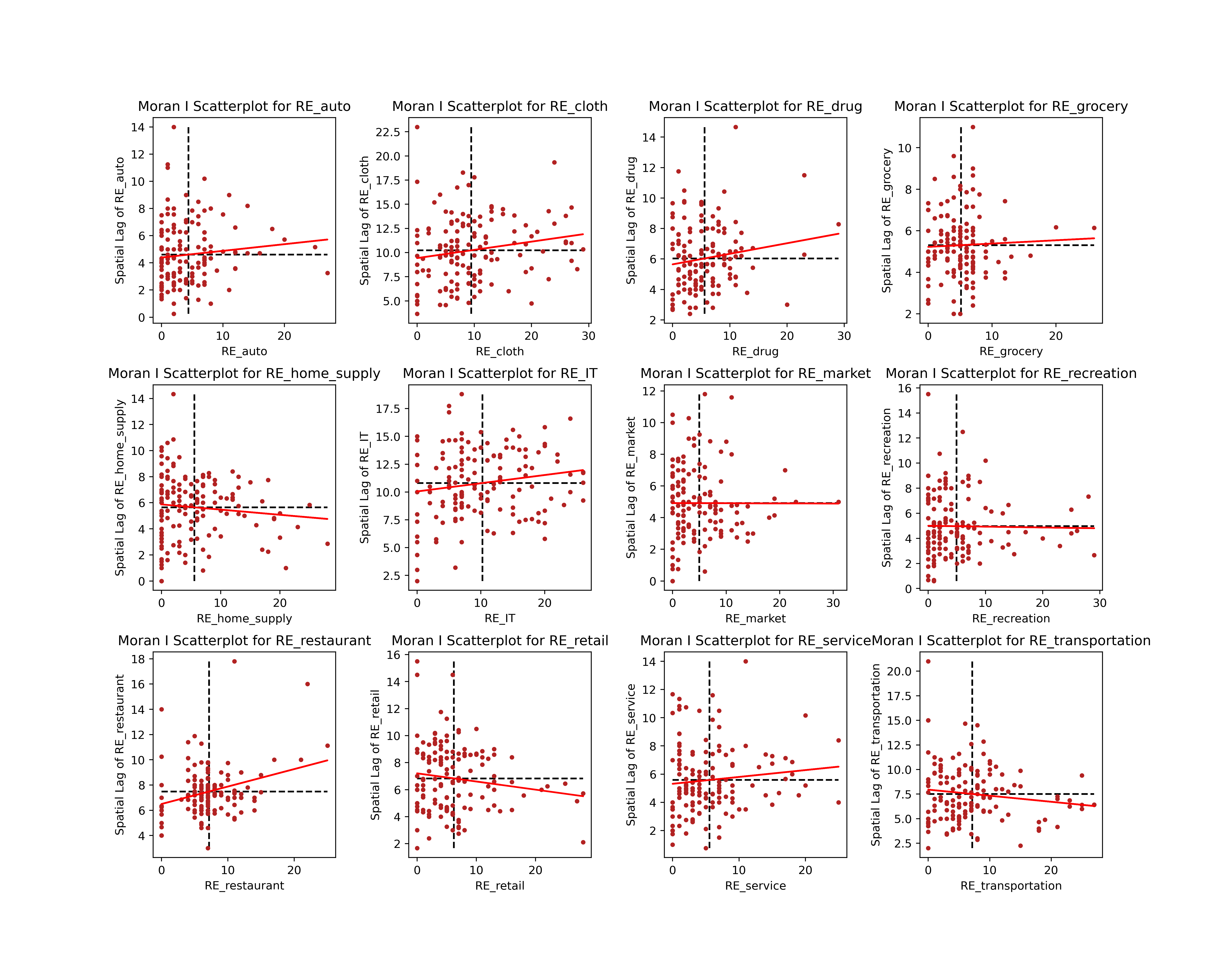}
\caption{Global Moran’s I scatterplot for recovery duration (RD: recovery duration)}
\end{figure}

\end{document}